\documentclass[%
 aip,
 amsmath,amssymb,
 reprint,%
]{revtex4-1}

\usepackage{graphicx}
\usepackage{dcolumn}
\usepackage{bm}
\usepackage{braket}

\usepackage{subcaption} 
\usepackage{siunitx} 

\usepackage[utf8]{inputenc}
\usepackage[T1]{fontenc}
\usepackage{mathptmx}

\usepackage{xcolor}

\begin{document}


\title[Multivariate Discrimination in Quantum Target Detection]{Multivariate Discrimination in Quantum Target Detection}

\author{Peter Svihra}
\email{peter.svihra@manchester.ac.uk}
\affiliation{Department of Physics, Faculty of Nuclear Sciences and Physical Engineering, Czech Technical University, Prague 115 19, Czech Republic}
\affiliation{Department of Physics and Astronomy, School of Natural Sciences, 
University of Manchester, Manchester M13 9PL, United Kingdom}

\author{Yingwen Zhang}
\author{Paul Hockett}
\affiliation{National Research Council of Canada, 100 Sussex Drive, Ottawa, Ontario, K1A 0R6, Canada}
\author{Steven Ferrante}
\affiliation{Physics Department, Brookhaven National Laboratory, Upton, NY 11973, USA}

\author{Benjamin Sussman}
\affiliation{National Research Council of Canada, 100 Sussex Drive, Ottawa, Ontario, K1A 0R6, Canada}\affiliation{Department of Physics, University of Ottawa, Ottawa, Ontario, K1N 6N5, Canada}

\author{Duncan England}
\affiliation{National Research Council of Canada, 100 Sussex Drive, Ottawa, Ontario, K1A 0R6, Canada}
\author{Andrei Nomerotski}
\affiliation{Physics Department, Brookhaven National Laboratory, Upton, NY 11973, USA}

\date{\today}

\begin{abstract}
We describe a simple multivariate technique of likelihood ratios for improved discrimination of signal and background in multi-dimensional quantum target detection. The technique combines two independent variables, time difference and summed energy, of a photon pair from the spontaneous parametric down-conversion source into an optimal discriminant. The discriminant performance was studied in experimental data and in Monte-Carlo modelling with clear improvement shown compared to previous techniques. As novel detectors become available, we expect this type of multivariate analysis to become increasingly important in multi-dimensional quantum optics.  
\end{abstract}

\maketitle

%
Non-classical correlation is at the heart of a range of quantum-enhanced technologies~\cite{Giovannetti2004,Ekert91,Genovese2016,Preskill2018}. In quantum optics, correlated photon pairs are routinely produced using the workhorse spontaneous parametric down-conversion (SPDC) sources. These sources have been used to generate pairs of photons that are correlated in almost every imaginable degree of freedom (DOF), including time~\cite{Marcikic2002}, polarization~\cite{Kwiat1999}, position-momentum~\cite{Howell2004}, orbital angular momentum (OAM)~\cite{Mair2001}, or frequency~\cite{Olislager2010}. The polarization degree of freedom naturally lends itself to quantum information theory, indeed, much of the seminal work in the field employed polarization-entangled photon pairs~\cite{Aspect1981,Genovese2005,Shalm2015,Yin2017}. However, polarization is by nature only two dimensional so each photon can carry only a single bit of information. Other DOFs are in principle unbounded offering high-dimensional encoding. But, while these high-dimensional states offer great promise, measuring them efficiently remains a significant challenge.
Traditional avalanche single photon detectors offer excellent temporal resolution but are single-mode so require scanning techniques to measure continuous DOFs~\cite{Howell2004,Poh2007,Stutz2007,Just2013,Forbes2019}. Alternatively, single photon sensitive cameras can be employed~\cite{Jachura2015, spad, Fickler2013,Allevi2014}, but they suffer from low frame rate making continuous readout with good temporal resolution impossible. 

The Tpx3Cam is an optical camera based on a technology originating in the high-energy physics community that has been adapted for optical detection by bonding a fast readout chip to an optical sensor \cite{Nomerotski2019}. The resulting spatial resolution is comparable to intensified CCD or EMCCD cameras, but with a so-called data-driven readout, only pixels in which the readout exceeds a threshold are read out allowing continuous operation and efficient time stamping with nanosecond resolution. By appending an image intensifier, the Tpx3Cam can be made to detect single photons, bringing a paradigm shift in quantum imaging devices. We expect this sensor to have applications in a range of quantum and quantum-inspired sensing techniques including ghost-imaging\cite{Meyers2008,Ferri2010,Shapiro2012} and quantum illumination \cite{Lloyd2008,Pirandola2018}.

In a recent paper~\cite{quantumillumination2019}, we applied the Tpx3Cam to quantum target detection -- a simplified form of quantum illumination that does not require entanglement from a photon pair source, only correlation.
Pairs of photons were generated by SPDC with one photon from each pair (the `herald’) measured locally and the other (the `signal’) sent to a target, which is hidden in a large amount of background light. After interacting with the target, correlations between the scattered signal photons and the herald photons are measured. This technique provides improved background rejection compared to simply measuring the back-scattered signal because the signal and herald modes are perfectly correlated, whereas the background is uncorrelated~\cite{Lloyd2008,Lopaeva2013,England2019}. 

\begin{figure}
    \centering
    \includegraphics[width=0.48\linewidth]{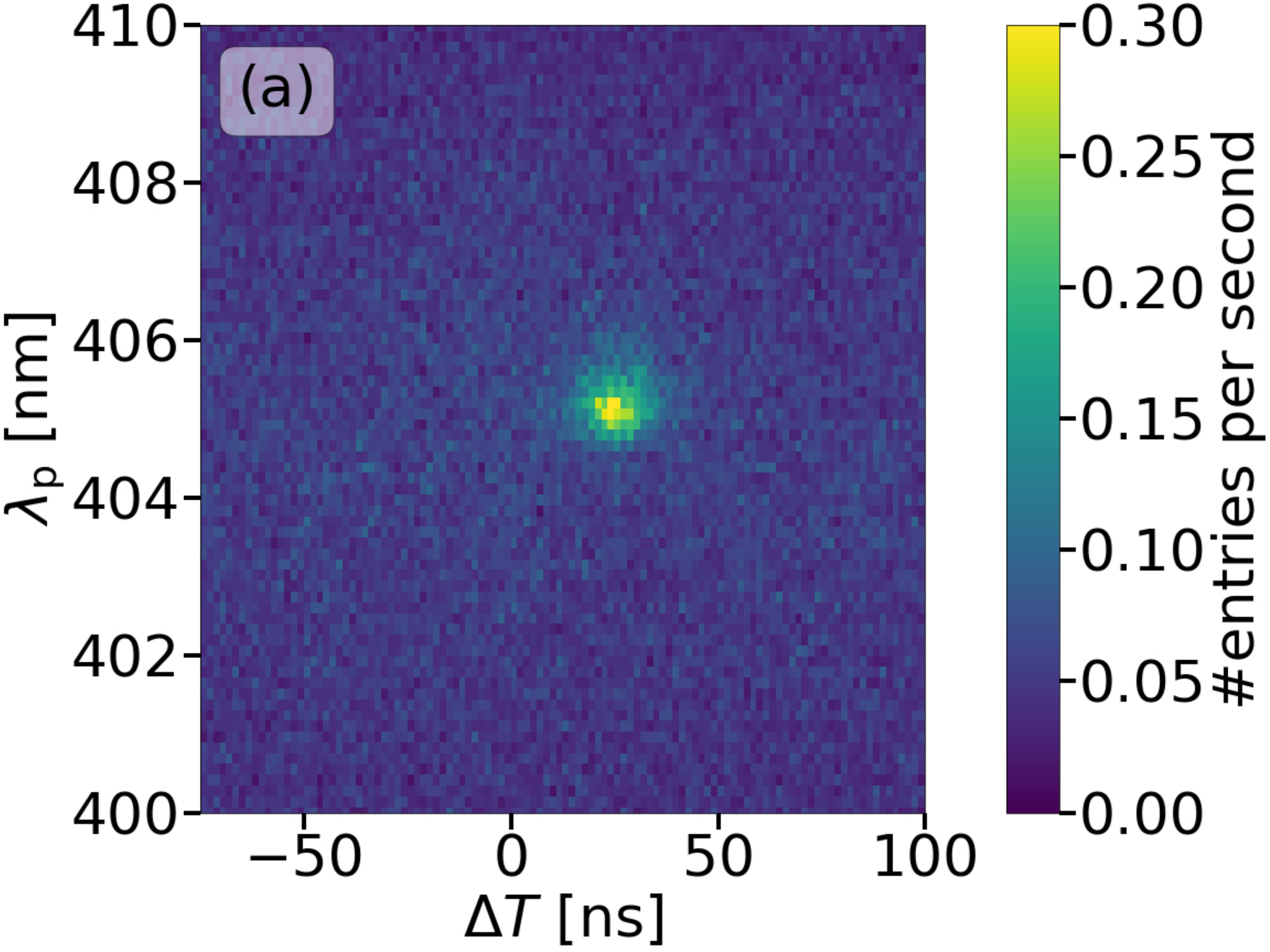}
    \hfil
    \includegraphics[width=0.48\linewidth]{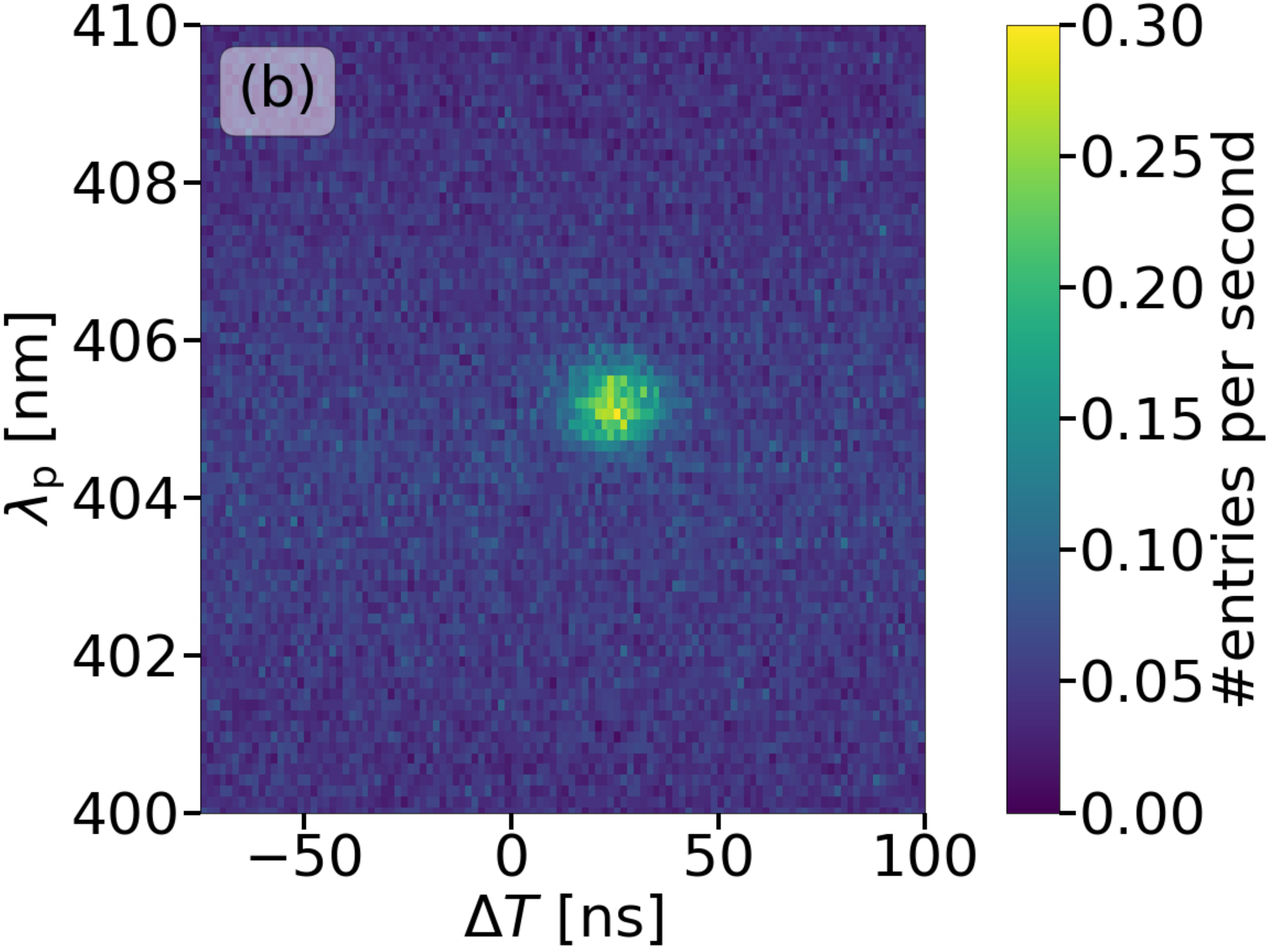}
    \caption{Two-dimensional distribution of spectroscopic and temporal variables for {\bf (a)} data and {\bf (b)} MC simulation..}
    \label{fig:2ddata}
\end{figure}

In previous work employing only timing correlations~\cite{England2019}, a peak in the one-dimensional histogram of photon arrival times reveals the presence of a target in the signal beam, and the timing delay of the peak gives the distance to the target~\cite{Liu2019}. The target can be said to be ‘detected’ if the size of the peak exceeds its statistical fluctuations by a predetermined amount (e.g. two-sigma). In photon-counting experiments, the statistical fluctuations scale as $1/\sqrt{n}$ where $n$ is the number of detected photons. So one must transmit enough photons to achieve the desired detection confidence.
By exploiting the multidimensional capabilities of the Tpx3Cam as a two-photon spectrometer, it was possible to simultaneously measure frequency and time correlations allowing us to generate a two-dimensional histogram of photon arrival time difference and frequency sum (see Figure~\ref{fig:2ddata}(a)). Because of the two-dimensional nature of the correlation, the background, i.e. accidental coincidences, is greatly reduced resulting in an increased detection confidence. Or, put another way, the same detection confidence can be achieved by sending fewer photons.

In our recent work~\cite{quantumillumination2019}, the multi-variable correlations were analysed in a simple fashion with a temporal `coincidence window’ used to isolate pairs of photons that arrive with the correct time separation and then, subsequently, a spectral cut selected appropriate frequency correlations.
In this work we show that the background can be further reduced, and therefore, detection confidence increased, by applying a multivariate, or combined, discriminant.
Optimal discrimination is widely used in particle physics~\cite{Lyons1986}, and other fields~\cite{Lim2010,Cash1979}, and in this work it has been applied to quantum optics. As the Tpx3Cam, and other readout-driven cameras~\cite{pimms1, pimms2, valerga2014}, become more prevalent in quantum optics we expect this type of analysis to become increasingly important beyond quantum target detection. Furthermore, it is simple to extend this analysis to higher dimensions, for example, to analyse multi-variable hyper-entangled states~\cite{Kwiat1997,Barreiro2005}.


Here we use one of the most straightforward multivariate techniques, likelihood ratio \cite{Lyons1986, Cowan1998, Albert84, anderson2003}, to combine the time difference and photon energy into a single discriminant. It can be shown that this combination is optimal, which means that the resulting discriminating variable provides the best possible background suppression at a given selection efficiency \cite{Neyman1933, Wilks1938}. For the below discussion it is also important that the two variables, time and energy, are independent i.e. the distribution of one is independent of any selection on the other \cite{BORISOV1998}. We also would like to emphasize that the experimental accuracy of the presented time-energy measurements are some orders of magnitude beyond the reach of time-energy entanglement effects and, therefore, we do not consider them.

Let us assume that there are $n$ variables, which have different distributions for signal and background. For independent variables the discriminant can be written as a product of ratios \cite{Cowan1998}:
\begin{align}
    Y = \frac{f^B(x_1, ..., x_n)}{f^S(x_1, ..., x_n)} = \displaystyle\prod_{i=1}^{n} \frac{f^B(x_i)}{f^S(x_i)} = \displaystyle\prod_{i=1}^{n} Y_i ,
    \label{eq1}
\end{align}
where $Y_i$ is the ratio of probability density functions for signal, $f_S$, and background, $f_B$. The above procedure is very simple and generalizes to any number of discriminating variables. 

The approach described above requires knowledge of the signal and background distributions for the variables which are used to form the discriminant. These distributions are measured experimentally and modeled using Monte-Carlo (MC) simulations.
The MC simulations allow us to test the discriminant and evaluate performance in different regimes that were not investigated experimentally (see supplementary information).

\begin{figure}
    \centering
    \includegraphics[width=1.0\linewidth]{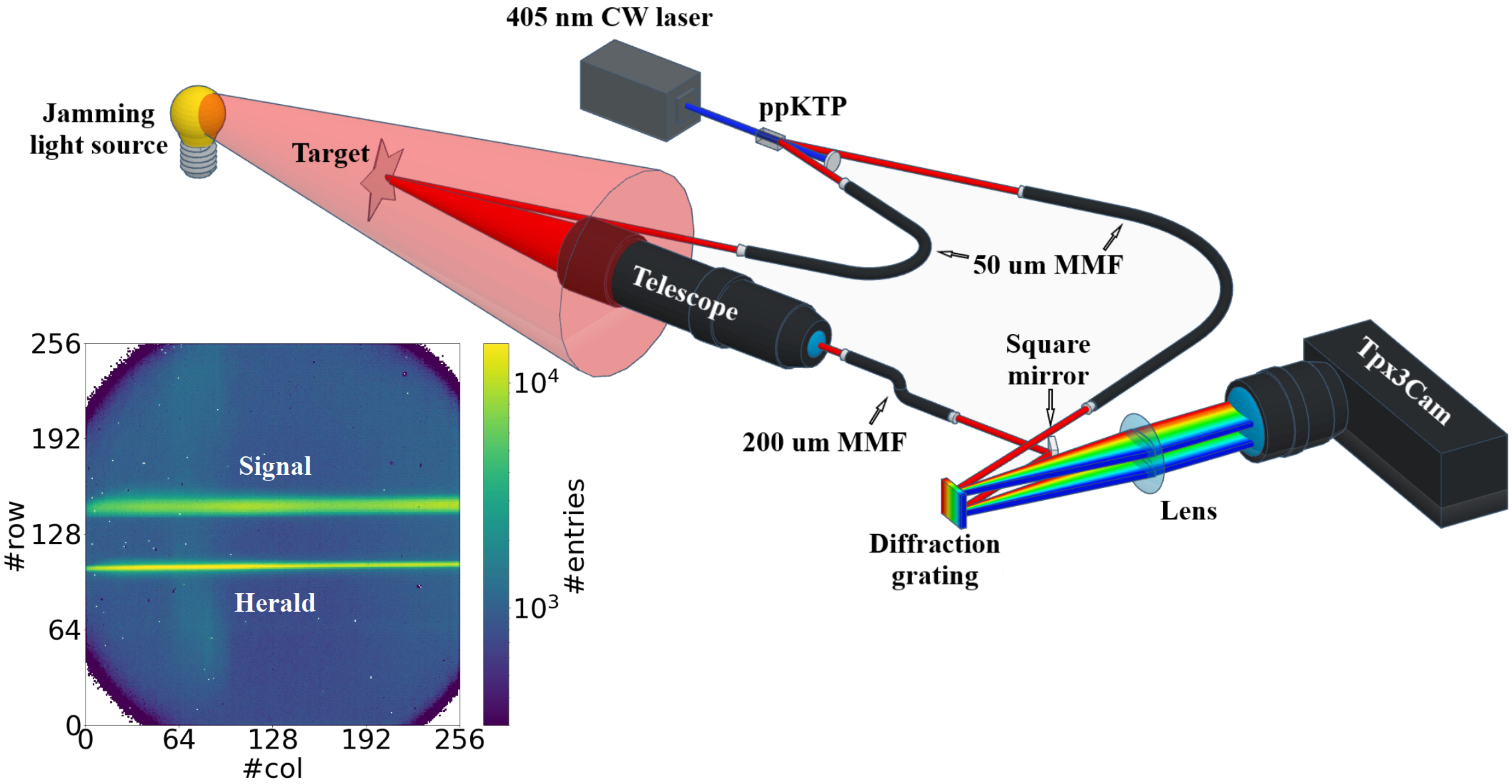}
    \caption{Schematic view of the experimental setup. 
    Inset: $x$, $y$ distribution of pixel occupancy in the camera. }
    \label{fig:setup}
\end{figure}

The experimental setup used for the measurements is shown schematically in Figure~\ref{fig:setup} and is described in detail elsewhere \cite{quantumillumination2019}. Briefly, an SPDC source is employed to produce pairs of photons with wavelength centered around  \SI{810}{\nano \meter}. One of the photons (signal) is sent onto a target and subsequently collected with a small telescope, while the other (herald) photon is sent directly to the camera. Before entering the fast camera the two photons are dispersed spectroscopically with a diffractive grating. The target is obscured by broadband `jamming' light from a halogen lamp introduced from behind the target.  

The fast camera, Tpx3Cam, is based on a Timepix3 chip \cite{timepix3} with \SI{1.5}{\nano \second} timing resolution coupled to an optical sensor \cite{timepixcam, Nomerotski2017}. The data obtained from the camera consists of $x$, $y$ position of hit pixels, ToA (Time of Arrival) and ToT (Time over Threshold) of the signal. The latter specifies deposited energy within the pixel. In order to achieve single photon sensitivity, an intensifier is employed, converting single photons to flashes of light, which are registered by the camera. The Hi-QE Red intensifier from Photonis \cite{Photonis} has a quantum efficiency of about 20\% at \SI{810}{\nano \meter} and employs P47 fast scintillator, which has timing performance compatible with nanosecond scale resolution \cite{P47}. All $256 \times 256$ pixels of the camera function independently with low dead-time and can be read out with a maximum total rate of about 10M photons per second \cite{spidr, ASI}. Similar configurations of the intensified Tpx3Cam have been used recently for characterization of quantum networks \cite{Ianzano2020, Nomerotski2020} and photon counting \cite{Nomerotski2020_1}.

The raw data is post-processed to identify `clusters', collections of pixels each corresponding to a single photon, and to perform centroiding. The centroiding improves the spatial resolution using a profile of the deposited energy in the cluster. We also apply a ToT-based correction to remove the time-walk effect in ToA and to further improve the time resolution. The post-processing steps are discussed in detail elsewhere \cite{Zhao2017, Ianzano2020}.


The inset of Figure~\ref{fig:setup} shows the measured data as a two-dimensional distribution of pixel occupancy of the camera data. The signal and herald modes after the diffractive grating appear as two horizontal stripes while the uniform background is mostly due to the intensifier dark counts and remaining stray light. In the spectrometer the photon wavelength has a linear relationship to the position along the stripe which can be derived by a simple calibration procedure \cite{quantumillumination2019}. The down-conversion process in the crystal requires conservation of energy and, therefore,
\begin{align}
    \frac{hc}{\lambda_\mathrm{p}} = \frac{hc}{\lambda_\mathrm{h}} + \frac{hc}{\lambda_\mathrm{s}},
\hspace{0.5 cm} \mathrm{implying}  \hspace{0.5 cm}
    \lambda_\mathrm{s} = \frac{\lambda_\mathrm{h}\lambda_\mathrm{p}}{\lambda_\mathrm{h} - \lambda_\mathrm{p}},
    \label{eq:signal}
\end{align}
where $\lambda_\mathrm{p}$ is the wavelength of the pump photon from the laser, \SI{405}{\nano \meter}, and $\lambda_\mathrm{h(s)}$ is the wavelength of the herald (signal) photon.
The spectral resolution is different for the herald and signal photons due to different style of multi-mode fibers used for their collection \cite{quantumillumination2019}, and is measured to be 1.6  and 3.2 pixels respectively for the herald and signal photons. The pump laser has a full-width half maximum linewidth of $\Delta \lambda_\mathrm{p} = \SI{0.6}{\nano \meter}$. 

To get the number of time coincidences for the photon pairs, we employed a previously used algorithm \cite{quantumillumination2019}. The data are selected according to the regions of interest, the two stripes, and for every event in one stripe, an event with the smallest $\Delta T$ ($\equiv \mathrm{ToA_1} - \mathrm{ToA_2}$) is found in the other stripe. 
The two-dimensional distribution of the sum energy and time difference of the photon pairs in the data is shown in Figure~\ref{fig:2ddata}(a). The sum energy, expressed through the pump photon wavelength, can be described by a normal distribution of width \SI{0.36}{\nano \meter} due to a combination of pump laser linewidth and spectrometer resolution. The time coincidence peak is also a normal distribution of width \SI{7.55}{\nano \second} due to the temporal resolution of the camera.

\begin{figure}
    \centering
     \includegraphics[width=1.0\linewidth]{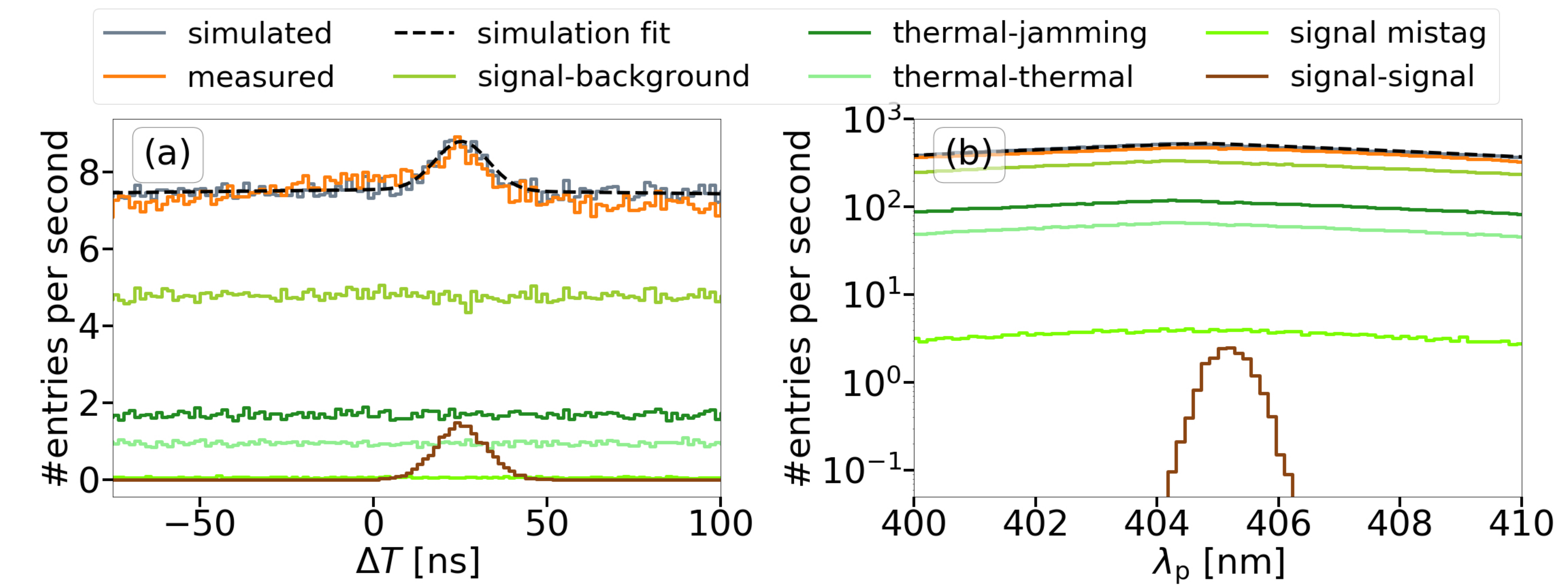}
     \caption{{\bf (a)} Time difference $\Delta T$ distributions for various sources of photon pairs. The true signal coincidence events are shown in the brown color histogram and different types of background events are shown in shades of green. “Signal mistag” are pairs of source photons originating from different pairs. “Thermal” refers to the DCR and residual stray photons. “Signal-background” includes all cases when a source photon is paired with a background photon. The fit function is a sum of a Gaussian (signal) and exponential (background). {\bf (b)} Pump photon wavelength $\lambda_\mathrm{p}$ distribution of the pairs. The fit function is the sum of a Gaussian (signal) and linear function (background).}
     \label{fig:deltaTDistribution}    
\end{figure}


To study the signal and background separation using the likelihood ratio discriminant we developed a MC model corresponding to experimental conditions such as signal and background resolutions and rates, including various inefficiencies of the whole system.
More details regarding the model and its matching to the dataset are described in the supplementary material.

\begin{figure}[t]
    \centering
    \includegraphics[width=0.48\linewidth]{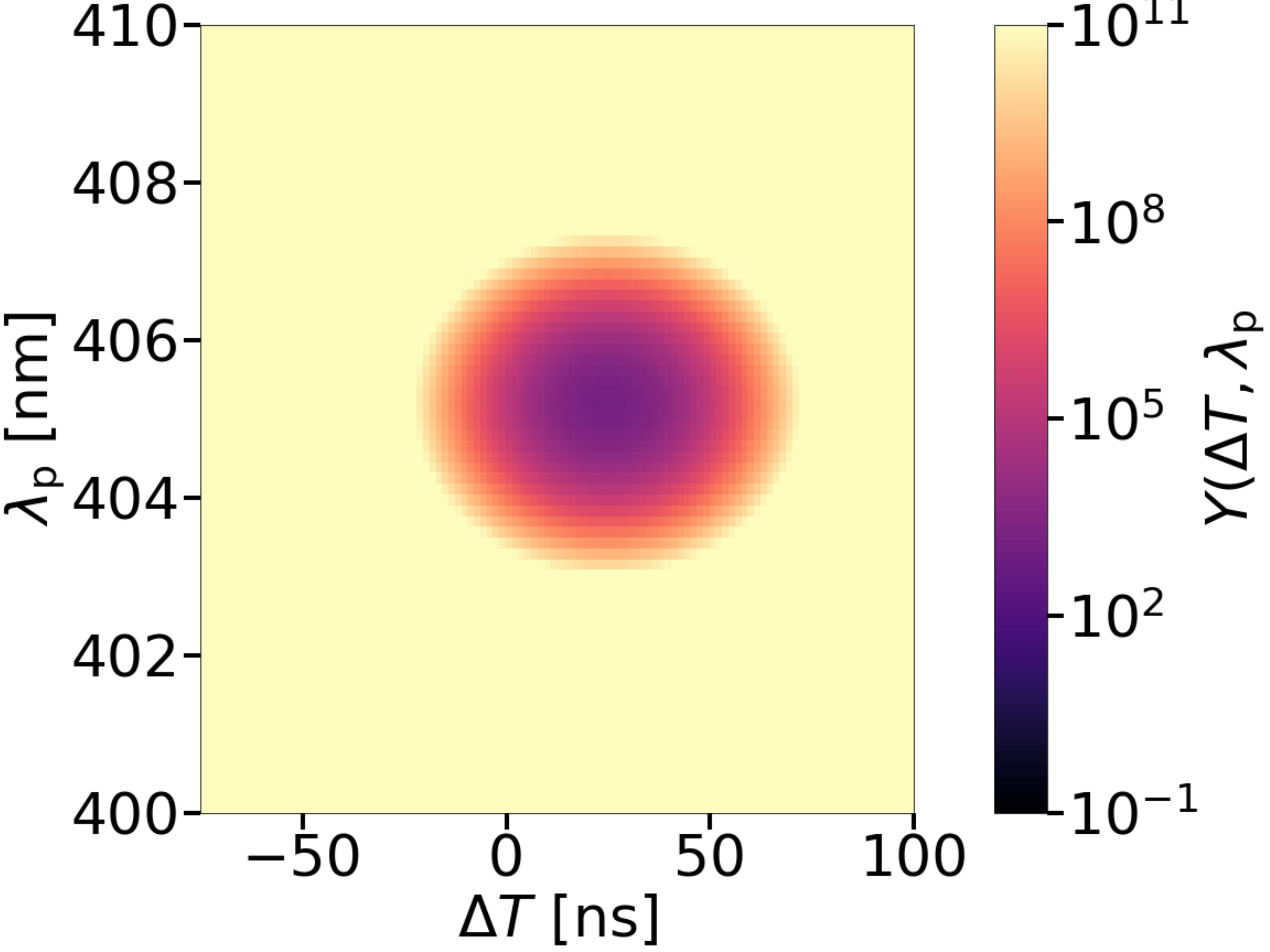}
    \caption{Two-dimensional likelihood ratio $Y$ for both time difference and pump photon wavelength.}
    \label{fig:ratios}
\end{figure}

Next we applied the aforementioned coincidence algorithm to find pairs of photons. It blindly processes the MC sample and determines which events are paired based on the closest ToA. We can then plot one-dimensional histograms of the  time difference $\Delta T$ distribution and the sum energy, represented by pump energy $\lambda_\mathrm{p}$, as plotted in Figure~\ref{fig:deltaTDistribution}(a) and (b), respectively. The MC simulations are in good agreement with the measured data. Note that each photon's origin was tagged in the MC to track them when forming photon pairs in the coincidence algorithm. This allows us to unambiguously find true signal coincidence events (brown color in Figure~\ref{fig:deltaTDistribution}), and identify different types of background events (shades of green). This is an useful feature of the MC simulation which is unavailable in experimental data. Figure~\ref{fig:deltaTDistribution} also illustrates very well that, before any selections are made, the signal to background ratio (SBR) is very poor.

The same data can be plotted in the two-dimensional representation shown in Figure~\ref{fig:2ddata} for both data and MC. The bright spot in the centre of the plot is due to true coincidences between photons produced in a pair which are highly correlated in time and anti-correlated in wavelength. The background is due to uncorrelated events such as photon-background or background-background coincidences. It is easy to see that signal-to-background contrast is far higher in the 2D representation than in either of the 1D histograms. Indeed, Figures~\ref{fig:2ddata} and~\ref{fig:deltaTDistribution} are a good visual representation of the difference between the `box-cuts' applied in our previous work, and the combined discriminant employed here. In the previous work, a region of interest was defined first in one degree of freedom (time), and then the other (energy) which is equivalent to selecting the peaks in the one dimensional histograms. Here, instead, the combined discriminant combines both variables when defining a region of interest effectively selecting an ellipse around the peak of the two-dimensional histogram. This idea is explored more rigorously below. 


Above, we studied in detail two discriminating variables, one derived from the temporal measurements and the other one derived from the spectroscopic measurements. We emphasize that this information is available on the pair by pair basis and, therefore, can be combined individually for each registered pair. The advantage of the utilized fast camera is in its ability to simultaneously record both: spatial coordinates as well as temporal information for each photon. 


The combined discriminant that we define is a function with two inputs: the photon pair time difference $\Delta T$ and wavelength of the reconstructed pump photon $\lambda_\mathrm{p}$. To combine $\Delta T$ and $\lambda_\mathrm{p}$, we start by defining background to signal ratios for these two variables,  $Y_\mathrm{\lambda_{p}}$ and $Y_\mathrm{\Delta T}$, as
\begin{align}
    Y_\mathrm{\lambda_{p}}(\lambda_\mathrm{p}) &= \frac{A |\lambda_\mathrm{p} - \lambda_\mathrm{b0}| + B} {\frac{N}{2\pi \sigma_{\lambda_\mathrm{p}}} \exp\left(-\frac{(\lambda_\mathrm{p} - \lambda_\mathrm{p0})^{2}}{2\sigma_{\lambda_{\mathrm{p}}}^{2}}\right)},  
    \label{eq:SBRenergy} \\
    Y_\mathrm{\Delta T}(\Delta T) &= \frac{C \exp\left(-\frac{|\Delta T|}{b}\right)}{\frac{N}{2\pi \sigma_{\Delta T_\mathrm{s}}} \exp\left(-\frac{(\Delta T-\Delta T_\mathrm{s0})^{2}}{2\sigma_{\Delta T_\mathrm{s}}^{2}}\right)}.
    \label{eq:SBRtime}
\end{align}

Where the probability density functions for the true coincidences (denominator of equations (\ref{eq:SBRenergy}) and (\ref{eq:SBRtime})) are assumed to be Gaussian in spectrum and time. The probability density function of the background noise (numerators of (\ref{eq:SBRenergy} and \ref{eq:SBRtime}) drop linearly and exponentially for the spectrum and time respectively. These are empirical models with all parameters tuned to fit experimental data (see Figure~\ref{fig:deltaTDistribution} and supplementary Figure~S2 ).

Combining the ratios $Y_\mathrm{\lambda_{p}}$ and $Y_\mathrm{\Delta T}$ according to (\ref{eq1}):
\begin{align}
    Y(\lambda_\mathrm{p}, \Delta T) \equiv Y(Y_{\lambda_\mathrm{p}}, Y_\mathrm{\Delta T}) = Y_{\lambda_\mathrm{p}} \cdot Y_\mathrm{\Delta T}
\end{align}
yields the two-dimensional likelihood ratio function $Y$ with the result shown in Figure~\ref{fig:ratios} with a deep, well-defined minimum of the function in the center.

\begin{figure}
    \centering
    \includegraphics[width=1.0\linewidth]{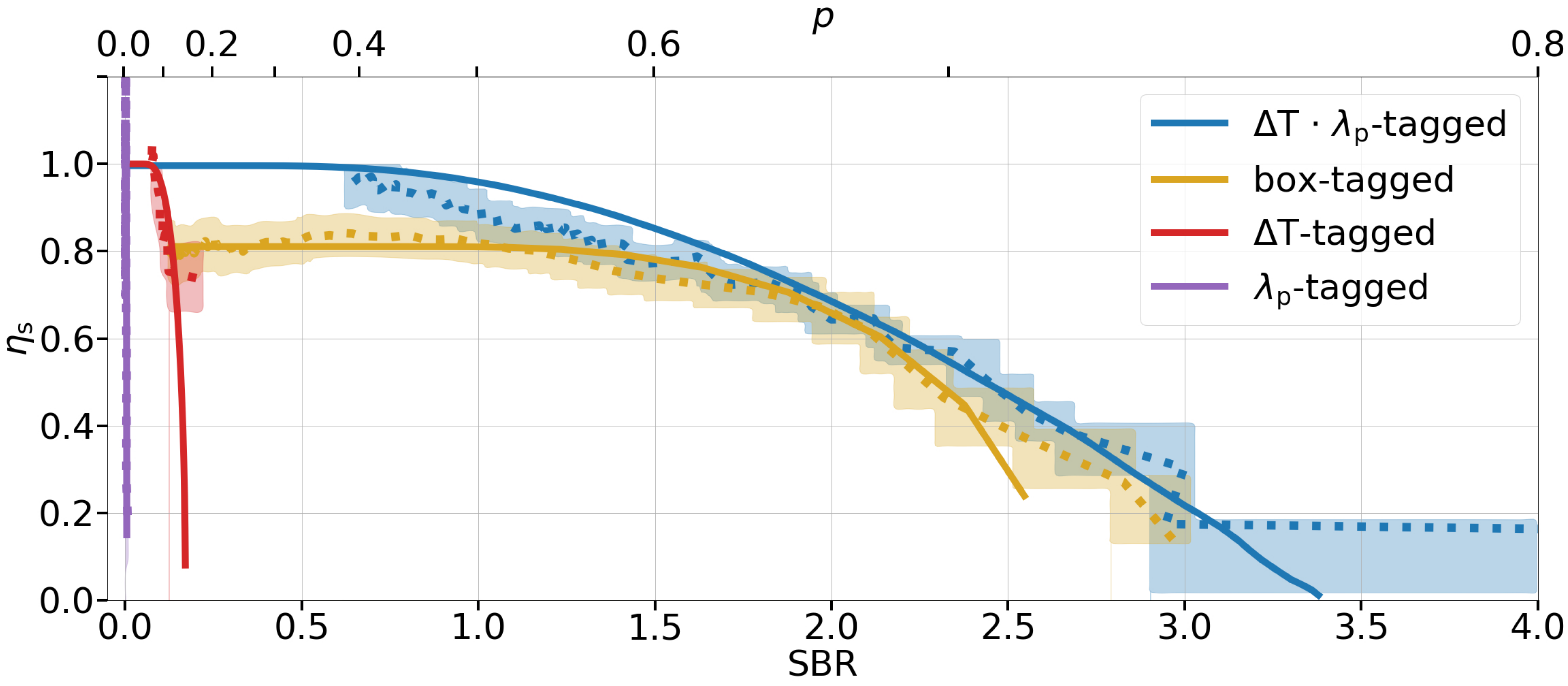}
    \caption{Selection efficiency $\eta_\mathrm{s}$ (ratio of signal counts to the total number of signal events) plotted as a function of sample purity $p$ for four cases: discriminants based on time difference, based on spectral information and based on the combined discriminant; and discriminant based on traditional box cuts, see the text. Solid lines are MC results, dotted lines are derived from the experimental data with errors shown as color bands.}
    \label{fig:precision}
\end{figure}

With the two-dimensional likelihood ratio calculated, we can proceed to use it to process the data. The selection criteria to eliminate background will be determined by drawing a Y-isoline around the peak in the 2D histogram; events outside this isoline are rejected and those inside are retained. Appropriate selection of the isoline is important, if it is too large then too many background events are included, reducing the SBR. If the isoline is too small, the SBR will be high, but signal events will be discarded along with the background. This trade-off is illustrated in Figure~\ref{fig:precision} where the SBR is plotted as a function of the selection efficiency $\eta_s$. Here we have defined the selection efficiency as the fraction of true coincidences that remain after selection. A figure of merit for this analysis is the SBR, defined simply as $SBR = s/b$, where $s$ is the signal counts and $b$ is the background. Alternatively we can consider the ‘sample purity’ $p$, which is a commonly used metric in multivariate analysis and is defined as $p=s/(s+b)$.


We tested the discriminating power of this newly obtained variable $Y$ by comparing it to $Y_{\lambda_\mathrm{p}}$ and  $Y_\mathrm{\Delta T}$ performances on their own. We also analysed the MC data using simple box-cuts where the temporal cut was fixed at $\pm10$\,ns and the spectral cut width was varied, as in reference~\cite{quantumillumination2019}.

In Figure~\ref{fig:precision}, the SBR, sample purity and selection efficiency are plotted for various selection parameters using different techniques: sum energy-only, time difference only, time-energy box cuts, and combined discriminant. In each case, as the selected region becomes smaller, the sample purity increases as more background is eliminated, but the selection efficiency decreases as true coincidences are also rejected. It is clear that the performance is vastly superior for the multivariate techniques compared to the single-variable approaches, and that the combined discriminant outperforms the box-cut method. According to the MC simulations, for a constant selection efficiency $\eta_\mathrm{s} = 0.80$ the SBR is increased by \SI{26}{\percent} for the optimal discriminant. For comparison, we also apply a similar analysis to the experimental data, shown with dotted lines in Figure~\ref{fig:precision}. The MC/data agreement is within errors, with the same functional form, confirming better performance of the optimal discriminant.

For target detection, the SBR is not the only relevant parameter, one must also consider the ratio of the number of counts to the statistical fluctuations, hereafter referred to as the signal-to-noise ratio (SNR). Unlike the SBR, which is constant, the SNR increases with longer integration time. Since the SNR includes statistical fluctuations in both the signal and the background, it is closely related to the SBR and an improvement in SBR will directly lead to an improvement in SNR~\cite{quantumillumination2019}. When using the optimal discriminant, an SBR improvement of 26\% corresponds to an SNR improvement of 7\%, for a given integration time. To put this another way, if target detection is defined to occur at a particular SNR threshold, then it will require 13\% fewer photons to detect a target using the optimal discriminant compared to the box cut.  


In summary, we employed a recently developed fast camera, Tpx3Cam, in the context of quantum target detection and considered an optimal discriminant based on the likelihood ratios for two measured variables, energy and time. The optimal discriminant can be established in advance through modelling of the experiment so tagging can be performed online with no computational overhead. We achieved a \SI{26}{\percent} improvement of SBR for the same selection efficiency compared to the previously used selections.

The optimal discriminant is also more resource efficient than previous techniques requiring 13\% fewer photons to achieve a detection threshold. We believe this multivariate approach is a promising avenue to analyse quantum sensing protocols using correlated photon pairs and, in general, high-dimensional quantum states. This analysis was performed on two-dimensional data but is easily extended to higher dimensions, we expect the improvements to be more pronounced when more variables are included. Finally, another opportunity for the future work is to extend the multivariate analysis to determination of distance to the target using all available information in the data. 
\newline

See supplementary material for more information on the MC model and on the performance predictions for different resolutions and background rates.

%
The authors are grateful to Philip Bustard, Denis Guay, and Doug Moffatt for technical support and stimulating discussion. We acknowledge support from Defence Research and Development Canada (DRDC) and from U.S. Department of Energy QuantISED award. S.F. acknowledges support under the Science Undergraduate Laboratory Internships Program (SULI) by the U.S. Department of Energy.
The work was also supported by the grant LM2018109 of Ministry of Education, Youth and Sports as well as by Centre of Advanced Applied Sciences CZ.02.1.01/0.0/0.0/16-019/0000778, co-financed by the European Union. 

The data that support the findings of this study are available from the corresponding author upon reasonable request.

\section*{References}
\bibliographystyle{aipnum4-1}
\bibliography{bibCDaip} 

\begin{thebibliography}{56}%
\makeatletter
\providecommand \@ifxundefined [1]{%
 \@ifx{#1\undefined}
}%
\providecommand \@ifnum [1]{%
 \ifnum #1\expandafter \@firstoftwo
 \else \expandafter \@secondoftwo
 \fi
}%
\providecommand \@ifx [1]{%
 \ifx #1\expandafter \@firstoftwo
 \else \expandafter \@secondoftwo
 \fi
}%
\providecommand \natexlab [1]{#1}%
\providecommand \enquote  [1]{``#1''}%
\providecommand \bibnamefont  [1]{#1}%
\providecommand \bibfnamefont [1]{#1}%
\providecommand \citenamefont [1]{#1}%
\providecommand \href@noop [0]{\@secondoftwo}%
\providecommand \href [0]{\begingroup \@sanitize@url \@href}%
\providecommand \@href[1]{\@@startlink{#1}\@@href}%
\providecommand \@@href[1]{\endgroup#1\@@endlink}%
\providecommand \@sanitize@url [0]{\catcode `\\12\catcode `\$12\catcode
  `\&12\catcode `\#12\catcode `\^12\catcode `\_12\catcode `\%12\relax}%
\providecommand \@@startlink[1]{}%
\providecommand \@@endlink[0]{}%
\providecommand \url  [0]{\begingroup\@sanitize@url \@url }%
\providecommand \@url [1]{\endgroup\@href {#1}{\urlprefix }}%
\providecommand \urlprefix  [0]{URL }%
\providecommand \Eprint [0]{\href }%
\providecommand \doibase [0]{http://dx.doi.org/}%
\providecommand \selectlanguage [0]{\@gobble}%
\providecommand \bibinfo  [0]{\@secondoftwo}%
\providecommand \bibfield  [0]{\@secondoftwo}%
\providecommand \translation [1]{[#1]}%
\providecommand \BibitemOpen [0]{}%
\providecommand \bibitemStop [0]{}%
\providecommand \bibitemNoStop [0]{.\EOS\space}%
\providecommand \EOS [0]{\spacefactor3000\relax}%
\providecommand \BibitemShut  [1]{\csname bibitem#1\endcsname}%
\let\auto@bib@innerbib\@empty
\bibitem [{\citenamefont {Giovannetti}, \citenamefont {Lloyd},\ and\
  \citenamefont {Maccone}(2004)}]{Giovannetti2004}%
  \BibitemOpen
  \bibfield  {author} {\bibinfo {author} {\bibfnamefont {V.}~\bibnamefont
  {Giovannetti}}, \bibinfo {author} {\bibfnamefont {S.}~\bibnamefont {Lloyd}},
  \ and\ \bibinfo {author} {\bibfnamefont {L.}~\bibnamefont {Maccone}},\ }\href
  {\doibase 10.1126/science.1104149} {\bibfield  {journal} {\bibinfo  {journal}
  {Science}\ }\textbf {\bibinfo {volume} {306}},\ \bibinfo {pages} {1330}
  (\bibinfo {year} {2004})}\BibitemShut {NoStop}%
\bibitem [{\citenamefont {Ekert}(1991)}]{Ekert91}%
  \BibitemOpen
  \bibfield  {author} {\bibinfo {author} {\bibfnamefont {A.~K.}\ \bibnamefont
  {Ekert}},\ }\href {\doibase 10.1103/PhysRevLett.67.661} {\bibfield  {journal}
  {\bibinfo  {journal} {Phys. Rev. Lett.}\ }\textbf {\bibinfo {volume} {67}},\
  \bibinfo {pages} {661} (\bibinfo {year} {1991})}\BibitemShut {NoStop}%
\bibitem [{\citenamefont {Genovese}(2016)}]{Genovese2016}%
  \BibitemOpen
  \bibfield  {author} {\bibinfo {author} {\bibfnamefont {M.}~\bibnamefont
  {Genovese}},\ }\href@noop {} {\bibfield  {journal} {\bibinfo  {journal}
  {Journal of Optics}\ }\textbf {\bibinfo {volume} {18}},\ \bibinfo {pages}
  {073002} (\bibinfo {year} {2016})}\BibitemShut {NoStop}%
\bibitem [{\citenamefont {Preskill}(2018)}]{Preskill2018}%
  \BibitemOpen
  \bibfield  {author} {\bibinfo {author} {\bibfnamefont {J.}~\bibnamefont
  {Preskill}},\ }\href {\doibase 10.22331/q-2018-08-06-79} {\bibfield
  {journal} {\bibinfo  {journal} {{Quantum}}\ }\textbf {\bibinfo {volume}
  {2}},\ \bibinfo {pages} {79} (\bibinfo {year} {2018})}\BibitemShut {NoStop}%
\bibitem [{\citenamefont {Marcikic}\ \emph {et~al.}(2002)\citenamefont
  {Marcikic}, \citenamefont {de~Riedmatten}, \citenamefont {Tittel},
  \citenamefont {Scarani}, \citenamefont {Zbinden},\ and\ \citenamefont
  {Gisin}}]{Marcikic2002}%
  \BibitemOpen
  \bibfield  {author} {\bibinfo {author} {\bibfnamefont {I.}~\bibnamefont
  {Marcikic}}, \bibinfo {author} {\bibfnamefont {H.}~\bibnamefont
  {de~Riedmatten}}, \bibinfo {author} {\bibfnamefont {W.}~\bibnamefont
  {Tittel}}, \bibinfo {author} {\bibfnamefont {V.}~\bibnamefont {Scarani}},
  \bibinfo {author} {\bibfnamefont {H.}~\bibnamefont {Zbinden}}, \ and\
  \bibinfo {author} {\bibfnamefont {N.}~\bibnamefont {Gisin}},\ }\href
  {\doibase 10.1103/PhysRevA.66.062308} {\bibfield  {journal} {\bibinfo
  {journal} {Phys. Rev. A}\ }\textbf {\bibinfo {volume} {66}},\ \bibinfo
  {pages} {062308} (\bibinfo {year} {2002})}\BibitemShut {NoStop}%
\bibitem [{\citenamefont {Kwiat}\ \emph {et~al.}(1999)\citenamefont {Kwiat},
  \citenamefont {Waks}, \citenamefont {White}, \citenamefont {Appelbaum},\ and\
  \citenamefont {Eberhard}}]{Kwiat1999}%
  \BibitemOpen
  \bibfield  {author} {\bibinfo {author} {\bibfnamefont {P.~G.}\ \bibnamefont
  {Kwiat}}, \bibinfo {author} {\bibfnamefont {E.}~\bibnamefont {Waks}},
  \bibinfo {author} {\bibfnamefont {A.~G.}\ \bibnamefont {White}}, \bibinfo
  {author} {\bibfnamefont {I.}~\bibnamefont {Appelbaum}}, \ and\ \bibinfo
  {author} {\bibfnamefont {P.~H.}\ \bibnamefont {Eberhard}},\ }\href {\doibase
  10.1103/PhysRevA.60.R773} {\bibfield  {journal} {\bibinfo  {journal} {Phys.
  Rev. A}\ }\textbf {\bibinfo {volume} {60}},\ \bibinfo {pages} {R773}
  (\bibinfo {year} {1999})}\BibitemShut {NoStop}%
\bibitem [{\citenamefont {Howell}\ \emph {et~al.}(2004)\citenamefont {Howell},
  \citenamefont {Bennink}, \citenamefont {Bentley},\ and\ \citenamefont
  {Boyd}}]{Howell2004}%
  \BibitemOpen
  \bibfield  {author} {\bibinfo {author} {\bibfnamefont {J.~C.}\ \bibnamefont
  {Howell}}, \bibinfo {author} {\bibfnamefont {R.~S.}\ \bibnamefont {Bennink}},
  \bibinfo {author} {\bibfnamefont {S.~J.}\ \bibnamefont {Bentley}}, \ and\
  \bibinfo {author} {\bibfnamefont {R.~W.}\ \bibnamefont {Boyd}},\ }\href
  {\doibase 10.1103/PhysRevLett.92.210403} {\bibfield  {journal} {\bibinfo
  {journal} {Phys. Rev. Lett.}\ }\textbf {\bibinfo {volume} {92}},\ \bibinfo
  {pages} {210403} (\bibinfo {year} {2004})}\BibitemShut {NoStop}%
\bibitem [{\citenamefont {Mair}\ \emph {et~al.}(2001)\citenamefont {Mair},
  \citenamefont {Vaziri}, \citenamefont {Weihs},\ and\ \citenamefont
  {Zeilinger}}]{Mair2001}%
  \BibitemOpen
  \bibfield  {author} {\bibinfo {author} {\bibfnamefont {A.}~\bibnamefont
  {Mair}}, \bibinfo {author} {\bibfnamefont {A.}~\bibnamefont {Vaziri}},
  \bibinfo {author} {\bibfnamefont {G.}~\bibnamefont {Weihs}}, \ and\ \bibinfo
  {author} {\bibfnamefont {A.}~\bibnamefont {Zeilinger}},\ }\href@noop {}
  {\bibfield  {journal} {\bibinfo  {journal} {Nature}\ }\textbf {\bibinfo
  {volume} {412}},\ \bibinfo {pages} {313} (\bibinfo {year}
  {2001})}\BibitemShut {NoStop}%
\bibitem [{\citenamefont {Olislager}\ \emph {et~al.}(2010)\citenamefont
  {Olislager}, \citenamefont {Cussey}, \citenamefont {Nguyen}, \citenamefont
  {Emplit}, \citenamefont {Massar}, \citenamefont {Merolla},\ and\
  \citenamefont {Huy}}]{Olislager2010}%
  \BibitemOpen
  \bibfield  {author} {\bibinfo {author} {\bibfnamefont {L.}~\bibnamefont
  {Olislager}}, \bibinfo {author} {\bibfnamefont {J.}~\bibnamefont {Cussey}},
  \bibinfo {author} {\bibfnamefont {A.~T.}\ \bibnamefont {Nguyen}}, \bibinfo
  {author} {\bibfnamefont {P.}~\bibnamefont {Emplit}}, \bibinfo {author}
  {\bibfnamefont {S.}~\bibnamefont {Massar}}, \bibinfo {author} {\bibfnamefont
  {J.-M.}\ \bibnamefont {Merolla}}, \ and\ \bibinfo {author} {\bibfnamefont
  {K.~P.}\ \bibnamefont {Huy}},\ }\href {\doibase 10.1103/PhysRevA.82.013804}
  {\bibfield  {journal} {\bibinfo  {journal} {Phys. Rev. A}\ }\textbf {\bibinfo
  {volume} {82}},\ \bibinfo {pages} {013804} (\bibinfo {year}
  {2010})}\BibitemShut {NoStop}%
\bibitem [{\citenamefont {Aspect}, \citenamefont {Grangier},\ and\
  \citenamefont {Roger}(1981)}]{Aspect1981}%
  \BibitemOpen
  \bibfield  {author} {\bibinfo {author} {\bibfnamefont {A.}~\bibnamefont
  {Aspect}}, \bibinfo {author} {\bibfnamefont {P.}~\bibnamefont {Grangier}}, \
  and\ \bibinfo {author} {\bibfnamefont {G.}~\bibnamefont {Roger}},\ }\href
  {\doibase 10.1103/PhysRevLett.47.460} {\bibfield  {journal} {\bibinfo
  {journal} {Phys. Rev. Lett.}\ }\textbf {\bibinfo {volume} {47}},\ \bibinfo
  {pages} {460} (\bibinfo {year} {1981})}\BibitemShut {NoStop}%
\bibitem [{\citenamefont {Genovese}(2005)}]{Genovese2005}%
  \BibitemOpen
  \bibfield  {author} {\bibinfo {author} {\bibfnamefont {M.}~\bibnamefont
  {Genovese}},\ }\href@noop {} {\bibfield  {journal} {\bibinfo  {journal}
  {Physics Reports}\ }\textbf {\bibinfo {volume} {413}},\ \bibinfo {pages} {319
  } (\bibinfo {year} {2005})}\BibitemShut {NoStop}%
\bibitem [{\citenamefont {Shalm}\ \emph {et~al.}(2015)\citenamefont {Shalm},
  \citenamefont {Meyer-Scott}, \citenamefont {Christensen}, \citenamefont
  {Bierhorst}, \citenamefont {Wayne}, \citenamefont {Stevens}, \citenamefont
  {Gerrits}, \citenamefont {Glancy}, \citenamefont {Hamel}, \citenamefont
  {Allman} \emph {et~al.}}]{Shalm2015}%
  \BibitemOpen
  \bibfield  {author} {\bibinfo {author} {\bibfnamefont {L.~K.}\ \bibnamefont
  {Shalm}}, \bibinfo {author} {\bibfnamefont {E.}~\bibnamefont {Meyer-Scott}},
  \bibinfo {author} {\bibfnamefont {B.~G.}\ \bibnamefont {Christensen}},
  \bibinfo {author} {\bibfnamefont {P.}~\bibnamefont {Bierhorst}}, \bibinfo
  {author} {\bibfnamefont {M.~A.}\ \bibnamefont {Wayne}}, \bibinfo {author}
  {\bibfnamefont {M.~J.}\ \bibnamefont {Stevens}}, \bibinfo {author}
  {\bibfnamefont {T.}~\bibnamefont {Gerrits}}, \bibinfo {author} {\bibfnamefont
  {S.}~\bibnamefont {Glancy}}, \bibinfo {author} {\bibfnamefont {D.~R.}\
  \bibnamefont {Hamel}}, \bibinfo {author} {\bibfnamefont {M.~S.}\ \bibnamefont
  {Allman}},  \emph {et~al.},\ }\href {\doibase 10.1103/PhysRevLett.115.250402}
  {\bibfield  {journal} {\bibinfo  {journal} {Phys. Rev. Lett.}\ }\textbf
  {\bibinfo {volume} {115}},\ \bibinfo {pages} {250402} (\bibinfo {year}
  {2015})}\BibitemShut {NoStop}%
\bibitem [{\citenamefont {Yin}\ \emph {et~al.}(2017)\citenamefont {Yin},
  \citenamefont {Cao}, \citenamefont {Li}, \citenamefont {Liao}, \citenamefont
  {Zhang}, \citenamefont {Ren}, \citenamefont {Cai}, \citenamefont {Liu},
  \citenamefont {Li}, \citenamefont {Dai} \emph {et~al.}}]{Yin2017}%
  \BibitemOpen
  \bibfield  {author} {\bibinfo {author} {\bibfnamefont {J.}~\bibnamefont
  {Yin}}, \bibinfo {author} {\bibfnamefont {Y.}~\bibnamefont {Cao}}, \bibinfo
  {author} {\bibfnamefont {Y.-H.}\ \bibnamefont {Li}}, \bibinfo {author}
  {\bibfnamefont {S.-K.}\ \bibnamefont {Liao}}, \bibinfo {author}
  {\bibfnamefont {L.}~\bibnamefont {Zhang}}, \bibinfo {author} {\bibfnamefont
  {J.-G.}\ \bibnamefont {Ren}}, \bibinfo {author} {\bibfnamefont {W.-Q.}\
  \bibnamefont {Cai}}, \bibinfo {author} {\bibfnamefont {W.-Y.}\ \bibnamefont
  {Liu}}, \bibinfo {author} {\bibfnamefont {B.}~\bibnamefont {Li}}, \bibinfo
  {author} {\bibfnamefont {H.}~\bibnamefont {Dai}},  \emph {et~al.},\ }\href
  {\doibase 10.1126/science.aan3211} {\bibfield  {journal} {\bibinfo  {journal}
  {Science}\ }\textbf {\bibinfo {volume} {356}},\ \bibinfo {pages} {1140}
  (\bibinfo {year} {2017})}\BibitemShut {NoStop}%
\bibitem [{\citenamefont {Poh}\ \emph {et~al.}(2007)\citenamefont {Poh},
  \citenamefont {Lum}, \citenamefont {Marcikic}, \citenamefont
  {Lamas-Linares},\ and\ \citenamefont {Kurtsiefer}}]{Poh2007}%
  \BibitemOpen
  \bibfield  {author} {\bibinfo {author} {\bibfnamefont {H.~S.}\ \bibnamefont
  {Poh}}, \bibinfo {author} {\bibfnamefont {C.~Y.}\ \bibnamefont {Lum}},
  \bibinfo {author} {\bibfnamefont {I.}~\bibnamefont {Marcikic}}, \bibinfo
  {author} {\bibfnamefont {A.}~\bibnamefont {Lamas-Linares}}, \ and\ \bibinfo
  {author} {\bibfnamefont {C.}~\bibnamefont {Kurtsiefer}},\ }\href {\doibase
  10.1103/PhysRevA.75.043816} {\bibfield  {journal} {\bibinfo  {journal} {Phys.
  Rev. A}\ }\textbf {\bibinfo {volume} {75}},\ \bibinfo {pages} {043816}
  (\bibinfo {year} {2007})}\BibitemShut {NoStop}%
\bibitem [{\citenamefont {Stütz}\ \emph {et~al.}(2007)\citenamefont {Stütz},
  \citenamefont {Gröblacher}, \citenamefont {Jennewein},\ and\ \citenamefont
  {Zeilinger}}]{Stutz2007}%
  \BibitemOpen
  \bibfield  {author} {\bibinfo {author} {\bibfnamefont {M.}~\bibnamefont
  {Stütz}}, \bibinfo {author} {\bibfnamefont {S.}~\bibnamefont {Gröblacher}},
  \bibinfo {author} {\bibfnamefont {T.}~\bibnamefont {Jennewein}}, \ and\
  \bibinfo {author} {\bibfnamefont {A.}~\bibnamefont {Zeilinger}},\ }\href
  {\doibase 10.1063/1.2752728} {\bibfield  {journal} {\bibinfo  {journal}
  {Applied Physics Letters}\ }\textbf {\bibinfo {volume} {90}},\ \bibinfo
  {pages} {261114} (\bibinfo {year} {2007})}\BibitemShut {NoStop}%
\bibitem [{\citenamefont {Just}\ \emph {et~al.}(2013)\citenamefont {Just},
  \citenamefont {Cavanna}, \citenamefont {Chekhova},\ and\ \citenamefont
  {Leuchs}}]{Just2013}%
  \BibitemOpen
  \bibfield  {author} {\bibinfo {author} {\bibfnamefont {F.}~\bibnamefont
  {Just}}, \bibinfo {author} {\bibfnamefont {A.}~\bibnamefont {Cavanna}},
  \bibinfo {author} {\bibfnamefont {M.~V.}\ \bibnamefont {Chekhova}}, \ and\
  \bibinfo {author} {\bibfnamefont {G.}~\bibnamefont {Leuchs}},\ }\href@noop {}
  {\bibfield  {journal} {\bibinfo  {journal} {New Journal of Physics}\ }\textbf
  {\bibinfo {volume} {15}},\ \bibinfo {pages} {083015} (\bibinfo {year}
  {2013})}\BibitemShut {NoStop}%
\bibitem [{\citenamefont {Forbes}\ and\ \citenamefont
  {Nape}(2019)}]{Forbes2019}%
  \BibitemOpen
  \bibfield  {author} {\bibinfo {author} {\bibfnamefont {A.}~\bibnamefont
  {Forbes}}\ and\ \bibinfo {author} {\bibfnamefont {I.}~\bibnamefont {Nape}},\
  }\href {\doibase 10.1116/1.5112027} {\bibfield  {journal} {\bibinfo
  {journal} {AVS Quantum Science}\ }\textbf {\bibinfo {volume} {1}},\ \bibinfo
  {pages} {011701} (\bibinfo {year} {2019})}\BibitemShut {NoStop}%
\bibitem [{\citenamefont {Jachura}\ and\ \citenamefont
  {Chrapkiewicz}(2015)}]{Jachura2015}%
  \BibitemOpen
  \bibfield  {author} {\bibinfo {author} {\bibfnamefont {M.}~\bibnamefont
  {Jachura}}\ and\ \bibinfo {author} {\bibfnamefont {R.}~\bibnamefont
  {Chrapkiewicz}},\ }\href {\doibase 10.1364/OL.40.001540} {\bibfield
  {journal} {\bibinfo  {journal} {Opt. Lett.}\ }\textbf {\bibinfo {volume}
  {40}},\ \bibinfo {pages} {1540} (\bibinfo {year} {2015})}\BibitemShut
  {NoStop}%
\bibitem [{\citenamefont {Gasparini}\ \emph {et~al.}(2017)\citenamefont
  {Gasparini}, \citenamefont {Bessire}, \citenamefont {Unternährer},
  \citenamefont {Stefanov}, \citenamefont {Boiko}, \citenamefont {Perenzoni},\
  and\ \citenamefont {Stoppa}}]{spad}%
  \BibitemOpen
  \bibfield  {author} {\bibinfo {author} {\bibfnamefont {L.}~\bibnamefont
  {Gasparini}}, \bibinfo {author} {\bibfnamefont {B.}~\bibnamefont {Bessire}},
  \bibinfo {author} {\bibfnamefont {M.}~\bibnamefont {Unternährer}}, \bibinfo
  {author} {\bibfnamefont {A.}~\bibnamefont {Stefanov}}, \bibinfo {author}
  {\bibfnamefont {D.}~\bibnamefont {Boiko}}, \bibinfo {author} {\bibfnamefont
  {M.}~\bibnamefont {Perenzoni}}, \ and\ \bibinfo {author} {\bibfnamefont
  {D.}~\bibnamefont {Stoppa}},\ }\href {\doibase 10.1117/12.2253598} {\bibfield
   {journal} {\bibinfo  {journal} {Proc.SPIE}\ }\textbf {\bibinfo {volume}
  {10111}},\ \bibinfo {pages} {10111 } (\bibinfo {year} {2017})}\BibitemShut
  {NoStop}%
\bibitem [{\citenamefont {Fickler}\ \emph {et~al.}(2013)\citenamefont
  {Fickler}, \citenamefont {Krenn}, \citenamefont {Lapkiewicz}, \citenamefont
  {Ramelow},\ and\ \citenamefont {Zeilinger}}]{Fickler2013}%
  \BibitemOpen
  \bibfield  {author} {\bibinfo {author} {\bibfnamefont {R.}~\bibnamefont
  {Fickler}}, \bibinfo {author} {\bibfnamefont {M.}~\bibnamefont {Krenn}},
  \bibinfo {author} {\bibfnamefont {R.}~\bibnamefont {Lapkiewicz}}, \bibinfo
  {author} {\bibfnamefont {S.}~\bibnamefont {Ramelow}}, \ and\ \bibinfo
  {author} {\bibfnamefont {A.}~\bibnamefont {Zeilinger}},\ }\href
  {http://dx.doi.org/10.1038/srep01914} {\bibfield  {journal} {\bibinfo
  {journal} {Scientific Reports}\ }\textbf {\bibinfo {volume} {3}},\ \bibinfo
  {pages} {1914} (\bibinfo {year} {2013})}\BibitemShut {NoStop}%
\bibitem [{\citenamefont {Allevi}\ \emph {et~al.}(2014)\citenamefont {Allevi},
  \citenamefont {Jedrkiewicz}, \citenamefont {Brambilla}, \citenamefont
  {Gatti}, \citenamefont {Pe\ifmmode~\check{r}\else \v{r}\fi{}ina},
  \citenamefont {Haderka},\ and\ \citenamefont {Bondani}}]{Allevi2014}%
  \BibitemOpen
  \bibfield  {author} {\bibinfo {author} {\bibfnamefont {A.}~\bibnamefont
  {Allevi}}, \bibinfo {author} {\bibfnamefont {O.}~\bibnamefont {Jedrkiewicz}},
  \bibinfo {author} {\bibfnamefont {E.}~\bibnamefont {Brambilla}}, \bibinfo
  {author} {\bibfnamefont {A.}~\bibnamefont {Gatti}}, \bibinfo {author}
  {\bibfnamefont {J.}~\bibnamefont {Pe\ifmmode~\check{r}\else \v{r}\fi{}ina}},
  \bibinfo {author} {\bibfnamefont {O.}~\bibnamefont {Haderka}}, \ and\
  \bibinfo {author} {\bibfnamefont {M.}~\bibnamefont {Bondani}},\ }\href@noop
  {} {\bibfield  {journal} {\bibinfo  {journal} {Phys. Rev. A}\ }\textbf
  {\bibinfo {volume} {90}},\ \bibinfo {pages} {063812} (\bibinfo {year}
  {2014})}\BibitemShut {NoStop}%
\bibitem [{\citenamefont {Nomerotski}(2019)}]{Nomerotski2019}%
  \BibitemOpen
  \bibfield  {author} {\bibinfo {author} {\bibfnamefont {A.}~\bibnamefont
  {Nomerotski}},\ }\href {\doibase https://doi.org/10.1016/j.nima.2019.05.034}
  {\bibfield  {journal} {\bibinfo  {journal} {Nuclear Instruments and Methods
  in Physics Research Section A: Accelerators, Spectrometers, Detectors and
  Associated Equipment}\ }\textbf {\bibinfo {volume} {937}},\ \bibinfo {pages}
  {26 } (\bibinfo {year} {2019})}\BibitemShut {NoStop}%
\bibitem [{\citenamefont {Meyers}, \citenamefont {Deacon},\ and\ \citenamefont
  {Shih}(2008)}]{Meyers2008}%
  \BibitemOpen
  \bibfield  {author} {\bibinfo {author} {\bibfnamefont {R.}~\bibnamefont
  {Meyers}}, \bibinfo {author} {\bibfnamefont {K.~S.}\ \bibnamefont {Deacon}},
  \ and\ \bibinfo {author} {\bibfnamefont {Y.}~\bibnamefont {Shih}},\
  }\href@noop {} {\bibfield  {journal} {\bibinfo  {journal} {Phys. Rev. A}\
  }\textbf {\bibinfo {volume} {77}},\ \bibinfo {pages} {041801} (\bibinfo
  {year} {2008})}\BibitemShut {NoStop}%
\bibitem [{\citenamefont {Ferri}\ \emph {et~al.}(2010)\citenamefont {Ferri},
  \citenamefont {Magatti}, \citenamefont {Lugiato},\ and\ \citenamefont
  {Gatti}}]{Ferri2010}%
  \BibitemOpen
  \bibfield  {author} {\bibinfo {author} {\bibfnamefont {F.}~\bibnamefont
  {Ferri}}, \bibinfo {author} {\bibfnamefont {D.}~\bibnamefont {Magatti}},
  \bibinfo {author} {\bibfnamefont {L.~A.}\ \bibnamefont {Lugiato}}, \ and\
  \bibinfo {author} {\bibfnamefont {A.}~\bibnamefont {Gatti}},\ }\href
  {\doibase 10.1103/PhysRevLett.104.253603} {\bibfield  {journal} {\bibinfo
  {journal} {Phys. Rev. Lett.}\ }\textbf {\bibinfo {volume} {104}},\ \bibinfo
  {pages} {253603} (\bibinfo {year} {2010})}\BibitemShut {NoStop}%
\bibitem [{\citenamefont {Shapiro}\ and\ \citenamefont
  {Boyd}(2012)}]{Shapiro2012}%
  \BibitemOpen
  \bibfield  {author} {\bibinfo {author} {\bibfnamefont {J.~H.}\ \bibnamefont
  {Shapiro}}\ and\ \bibinfo {author} {\bibfnamefont {R.~W.}\ \bibnamefont
  {Boyd}},\ }\href@noop {} {\bibfield  {journal} {\bibinfo  {journal} {Quantum
  Information Processing}\ }\textbf {\bibinfo {volume} {11}},\ \bibinfo {pages}
  {949} (\bibinfo {year} {2012})}\BibitemShut {NoStop}%
\bibitem [{\citenamefont {Lloyd}(2008)}]{Lloyd2008}%
  \BibitemOpen
  \bibfield  {author} {\bibinfo {author} {\bibfnamefont {S.}~\bibnamefont
  {Lloyd}},\ }\href {\doibase 10.1126/science.1160627} {\bibfield  {journal}
  {\bibinfo  {journal} {Science}\ }\textbf {\bibinfo {volume} {321}},\ \bibinfo
  {pages} {1463} (\bibinfo {year} {2008})}\BibitemShut {NoStop}%
\bibitem [{\citenamefont {Pirandola}\ \emph {et~al.}(2018)\citenamefont
  {Pirandola}, \citenamefont {Bardhan}, \citenamefont {Gehring}, \citenamefont
  {Weedbrook},\ and\ \citenamefont {Lloyd}}]{Pirandola2018}%
  \BibitemOpen
  \bibfield  {author} {\bibinfo {author} {\bibfnamefont {S.}~\bibnamefont
  {Pirandola}}, \bibinfo {author} {\bibfnamefont {B.~R.}\ \bibnamefont
  {Bardhan}}, \bibinfo {author} {\bibfnamefont {T.}~\bibnamefont {Gehring}},
  \bibinfo {author} {\bibfnamefont {C.}~\bibnamefont {Weedbrook}}, \ and\
  \bibinfo {author} {\bibfnamefont {S.}~\bibnamefont {Lloyd}},\ }\href@noop {}
  {\bibfield  {journal} {\bibinfo  {journal} {Nature Photonics}\ }\textbf
  {\bibinfo {volume} {12}},\ \bibinfo {pages} {724} (\bibinfo {year}
  {2018})}\BibitemShut {NoStop}%
\bibitem [{\citenamefont {Zhang}\ \emph {et~al.}(2020)\citenamefont {Zhang},
  \citenamefont {England}, \citenamefont {Nomerotski}, \citenamefont {Svihra},
  \citenamefont {Ferrante}, \citenamefont {Hockett},\ and\ \citenamefont
  {Sussman}}]{quantumillumination2019}%
  \BibitemOpen
  \bibfield  {author} {\bibinfo {author} {\bibfnamefont {Y.}~\bibnamefont
  {Zhang}}, \bibinfo {author} {\bibfnamefont {D.}~\bibnamefont {England}},
  \bibinfo {author} {\bibfnamefont {A.}~\bibnamefont {Nomerotski}}, \bibinfo
  {author} {\bibfnamefont {P.}~\bibnamefont {Svihra}}, \bibinfo {author}
  {\bibfnamefont {S.}~\bibnamefont {Ferrante}}, \bibinfo {author}
  {\bibfnamefont {P.}~\bibnamefont {Hockett}}, \ and\ \bibinfo {author}
  {\bibfnamefont {B.}~\bibnamefont {Sussman}},\ }\href {\doibase
  10.1103/PhysRevA.101.053808} {\bibfield  {journal} {\bibinfo  {journal}
  {Phys. Rev. A}\ }\textbf {\bibinfo {volume} {101}},\ \bibinfo {pages}
  {053808} (\bibinfo {year} {2020})}\BibitemShut {NoStop}%
\bibitem [{\citenamefont {Lopaeva}\ \emph {et~al.}(2013)\citenamefont
  {Lopaeva}, \citenamefont {Ruo~Berchera}, \citenamefont {Degiovanni},
  \citenamefont {Olivares}, \citenamefont {Brida},\ and\ \citenamefont
  {Genovese}}]{Lopaeva2013}%
  \BibitemOpen
  \bibfield  {author} {\bibinfo {author} {\bibfnamefont {E.~D.}\ \bibnamefont
  {Lopaeva}}, \bibinfo {author} {\bibfnamefont {I.}~\bibnamefont
  {Ruo~Berchera}}, \bibinfo {author} {\bibfnamefont {I.~P.}\ \bibnamefont
  {Degiovanni}}, \bibinfo {author} {\bibfnamefont {S.}~\bibnamefont
  {Olivares}}, \bibinfo {author} {\bibfnamefont {G.}~\bibnamefont {Brida}}, \
  and\ \bibinfo {author} {\bibfnamefont {M.}~\bibnamefont {Genovese}},\ }\href
  {\doibase 10.1103/PhysRevLett.110.153603} {\bibfield  {journal} {\bibinfo
  {journal} {Phys. Rev. Lett.}\ }\textbf {\bibinfo {volume} {110}},\ \bibinfo
  {pages} {153603} (\bibinfo {year} {2013})}\BibitemShut {NoStop}%
\bibitem [{\citenamefont {England}, \citenamefont {Balaji},\ and\ \citenamefont
  {Sussman}(2019)}]{England2019}%
  \BibitemOpen
  \bibfield  {author} {\bibinfo {author} {\bibfnamefont {D.~G.}\ \bibnamefont
  {England}}, \bibinfo {author} {\bibfnamefont {B.}~\bibnamefont {Balaji}}, \
  and\ \bibinfo {author} {\bibfnamefont {B.~J.}\ \bibnamefont {Sussman}},\
  }\href {\doibase 10.1103/PhysRevA.99.023828} {\bibfield  {journal} {\bibinfo
  {journal} {Phys. Rev. A}\ }\textbf {\bibinfo {volume} {99}},\ \bibinfo
  {pages} {023828} (\bibinfo {year} {2019})}\BibitemShut {NoStop}%
\bibitem [{\citenamefont {Liu}\ \emph {et~al.}(2019)\citenamefont {Liu},
  \citenamefont {Giovannini}, \citenamefont {He}, \citenamefont {England},
  \citenamefont {Sussman}, \citenamefont {Balaji},\ and\ \citenamefont
  {Helmy}}]{Liu2019}%
  \BibitemOpen
  \bibfield  {author} {\bibinfo {author} {\bibfnamefont {H.}~\bibnamefont
  {Liu}}, \bibinfo {author} {\bibfnamefont {D.}~\bibnamefont {Giovannini}},
  \bibinfo {author} {\bibfnamefont {H.}~\bibnamefont {He}}, \bibinfo {author}
  {\bibfnamefont {D.}~\bibnamefont {England}}, \bibinfo {author} {\bibfnamefont
  {B.~J.}\ \bibnamefont {Sussman}}, \bibinfo {author} {\bibfnamefont
  {B.}~\bibnamefont {Balaji}}, \ and\ \bibinfo {author} {\bibfnamefont {A.~S.}\
  \bibnamefont {Helmy}},\ }\href {\doibase 10.1364/OPTICA.6.001349} {\bibfield
  {journal} {\bibinfo  {journal} {Optica}\ }\textbf {\bibinfo {volume} {6}},\
  \bibinfo {pages} {1349} (\bibinfo {year} {2019})}\BibitemShut {NoStop}%
\bibitem [{\citenamefont {Lyons}(1986)}]{Lyons1986}%
  \BibitemOpen
  \bibfield  {author} {\bibinfo {author} {\bibfnamefont {L.}~\bibnamefont
  {Lyons}},\ }\href {\doibase 10.1017/CBO9781139167710} {\emph {\bibinfo
  {title} {Statistics for Nuclear and Particle Physicists}}}\ (\bibinfo
  {publisher} {Cambridge University Press},\ \bibinfo {year}
  {1986})\BibitemShut {NoStop}%
\bibitem [{\citenamefont {Lim}, \citenamefont {Li},\ and\ \citenamefont
  {Lee}(2010)}]{Lim2010}%
  \BibitemOpen
  \bibfield  {author} {\bibinfo {author} {\bibfnamefont {J.}~\bibnamefont
  {Lim}}, \bibinfo {author} {\bibfnamefont {E.}~\bibnamefont {Li}}, \ and\
  \bibinfo {author} {\bibfnamefont {S.-J.}\ \bibnamefont {Lee}},\ }\href
  {\doibase https://doi.org/10.1016/j.jmva.2009.10.011} {\bibfield  {journal}
  {\bibinfo  {journal} {Journal of Multivariate Analysis}\ }\textbf {\bibinfo
  {volume} {101}},\ \bibinfo {pages} {541 } (\bibinfo {year}
  {2010})}\BibitemShut {NoStop}%
\bibitem [{\citenamefont {{Cash}}(1979)}]{Cash1979}%
  \BibitemOpen
  \bibfield  {author} {\bibinfo {author} {\bibfnamefont {W.}~\bibnamefont
  {{Cash}}},\ }\href {\doibase 10.1086/156922} {\bibfield  {journal} {\bibinfo
  {journal} {\apj}\ }\textbf {\bibinfo {volume} {228}},\ \bibinfo {pages} {939}
  (\bibinfo {year} {1979})}\BibitemShut {NoStop}%
\bibitem [{\citenamefont {Nomerotski}\ \emph {et~al.}(2010)\citenamefont
  {Nomerotski}, \citenamefont {Brouard}, \citenamefont {Campbell},
  \citenamefont {Clark}, \citenamefont {Crooks}, \citenamefont {Fopma},
  \citenamefont {John}, \citenamefont {Johnsen}, \citenamefont {Slater},
  \citenamefont {Turchetta} \emph {et~al.}}]{pimms1}%
  \BibitemOpen
  \bibfield  {author} {\bibinfo {author} {\bibfnamefont {A.}~\bibnamefont
  {Nomerotski}}, \bibinfo {author} {\bibfnamefont {M.}~\bibnamefont {Brouard}},
  \bibinfo {author} {\bibfnamefont {E.}~\bibnamefont {Campbell}}, \bibinfo
  {author} {\bibfnamefont {A.}~\bibnamefont {Clark}}, \bibinfo {author}
  {\bibfnamefont {J.}~\bibnamefont {Crooks}}, \bibinfo {author} {\bibfnamefont
  {J.}~\bibnamefont {Fopma}}, \bibinfo {author} {\bibfnamefont {J.~J.}\
  \bibnamefont {John}}, \bibinfo {author} {\bibfnamefont {A.~J.}\ \bibnamefont
  {Johnsen}}, \bibinfo {author} {\bibfnamefont {C.}~\bibnamefont {Slater}},
  \bibinfo {author} {\bibfnamefont {R.}~\bibnamefont {Turchetta}},  \emph
  {et~al.},\ }\href@noop {} {\bibfield  {journal} {\bibinfo  {journal} {Journal
  of Instrumentation}\ }\textbf {\bibinfo {volume} {5}},\ \bibinfo {pages}
  {C07007} (\bibinfo {year} {2010})}\BibitemShut {NoStop}%
\bibitem [{\citenamefont {John}\ \emph {et~al.}(2012)\citenamefont {John},
  \citenamefont {Brouard}, \citenamefont {Clark}, \citenamefont {Crooks},
  \citenamefont {Halford}, \citenamefont {Hill}, \citenamefont {Lee},
  \citenamefont {Nomerotski}, \citenamefont {Pisarczyk}, \citenamefont
  {Sedgwick} \emph {et~al.}}]{pimms2}%
  \BibitemOpen
  \bibfield  {author} {\bibinfo {author} {\bibfnamefont {J.~J.}\ \bibnamefont
  {John}}, \bibinfo {author} {\bibfnamefont {M.}~\bibnamefont {Brouard}},
  \bibinfo {author} {\bibfnamefont {A.}~\bibnamefont {Clark}}, \bibinfo
  {author} {\bibfnamefont {J.}~\bibnamefont {Crooks}}, \bibinfo {author}
  {\bibfnamefont {E.}~\bibnamefont {Halford}}, \bibinfo {author} {\bibfnamefont
  {L.}~\bibnamefont {Hill}}, \bibinfo {author} {\bibfnamefont {J.~W.~L.}\
  \bibnamefont {Lee}}, \bibinfo {author} {\bibfnamefont {A.}~\bibnamefont
  {Nomerotski}}, \bibinfo {author} {\bibfnamefont {R.}~\bibnamefont
  {Pisarczyk}}, \bibinfo {author} {\bibfnamefont {I.}~\bibnamefont {Sedgwick}},
   \emph {et~al.},\ }\href@noop {} {\bibfield  {journal} {\bibinfo  {journal}
  {Journal of Instrumentation}\ }\textbf {\bibinfo {volume} {7}},\ \bibinfo
  {pages} {C08001} (\bibinfo {year} {2012})}\BibitemShut {NoStop}%
\bibitem [{\citenamefont {Vallerga}\ \emph {et~al.}(2014)\citenamefont
  {Vallerga}, \citenamefont {Tremsin}, \citenamefont {DeFazio}, \citenamefont
  {Michel}, \citenamefont {Alozy}, \citenamefont {Tick},\ and\ \citenamefont
  {Campbell}}]{valerga2014}%
  \BibitemOpen
  \bibfield  {author} {\bibinfo {author} {\bibfnamefont {J.}~\bibnamefont
  {Vallerga}}, \bibinfo {author} {\bibfnamefont {A.}~\bibnamefont {Tremsin}},
  \bibinfo {author} {\bibfnamefont {J.}~\bibnamefont {DeFazio}}, \bibinfo
  {author} {\bibfnamefont {T.}~\bibnamefont {Michel}}, \bibinfo {author}
  {\bibfnamefont {J.}~\bibnamefont {Alozy}}, \bibinfo {author} {\bibfnamefont
  {T.}~\bibnamefont {Tick}}, \ and\ \bibinfo {author} {\bibfnamefont
  {M.}~\bibnamefont {Campbell}},\ }\href@noop {} {\bibfield  {journal}
  {\bibinfo  {journal} {Journal of Instrumentation}\ }\textbf {\bibinfo
  {volume} {9}},\ \bibinfo {pages} {C05055} (\bibinfo {year}
  {2014})}\BibitemShut {NoStop}%
\bibitem [{\citenamefont {Kwiat}(1997)}]{Kwiat1997}%
  \BibitemOpen
  \bibfield  {author} {\bibinfo {author} {\bibfnamefont {P.~G.}\ \bibnamefont
  {Kwiat}},\ }\href@noop {} {\bibfield  {journal} {\bibinfo  {journal} {Journal
  of modern optics}\ }\textbf {\bibinfo {volume} {44}},\ \bibinfo {pages}
  {2173} (\bibinfo {year} {1997})}\BibitemShut {NoStop}%
\bibitem [{\citenamefont {Barreiro}\ \emph {et~al.}(2005)\citenamefont
  {Barreiro}, \citenamefont {Langford}, \citenamefont {Peters},\ and\
  \citenamefont {Kwiat}}]{Barreiro2005}%
  \BibitemOpen
  \bibfield  {author} {\bibinfo {author} {\bibfnamefont {J.~T.}\ \bibnamefont
  {Barreiro}}, \bibinfo {author} {\bibfnamefont {N.~K.}\ \bibnamefont
  {Langford}}, \bibinfo {author} {\bibfnamefont {N.~A.}\ \bibnamefont
  {Peters}}, \ and\ \bibinfo {author} {\bibfnamefont {P.~G.}\ \bibnamefont
  {Kwiat}},\ }\href {\doibase 10.1103/PhysRevLett.95.260501} {\bibfield
  {journal} {\bibinfo  {journal} {Phys. Rev. Lett.}\ }\textbf {\bibinfo
  {volume} {95}},\ \bibinfo {pages} {260501} (\bibinfo {year}
  {2005})}\BibitemShut {NoStop}%
\bibitem [{\citenamefont {Cowan}(1998)}]{Cowan1998}%
  \BibitemOpen
  \bibfield  {author} {\bibinfo {author} {\bibfnamefont {G.}~\bibnamefont
  {Cowan}},\ }\href {https://books.google.com/books?id=ff8ZyW0nlJAC} {\emph
  {\bibinfo {title} {Statistical Data Analysis}}},\ Oxford science
  publications\ (\bibinfo  {publisher} {Clarendon Press},\ \bibinfo {year}
  {1998})\BibitemShut {NoStop}%
\bibitem [{\citenamefont {Albert}\ and\ \citenamefont
  {Anderson}(1984)}]{Albert84}%
  \BibitemOpen
  \bibfield  {author} {\bibinfo {author} {\bibfnamefont {A.}~\bibnamefont
  {Albert}}\ and\ \bibinfo {author} {\bibfnamefont {J.~A.}\ \bibnamefont
  {Anderson}},\ }\href@noop {} {\bibfield  {journal} {\bibinfo  {journal}
  {Bka}\ }\textbf {\bibinfo {volume} {71}},\ \bibinfo {pages} {1} (\bibinfo
  {year} {1984})}\BibitemShut {NoStop}%
\bibitem [{\citenamefont {Anderson}(2003)}]{anderson2003}%
  \BibitemOpen
  \bibfield  {author} {\bibinfo {author} {\bibfnamefont {T.~W.}\ \bibnamefont
  {Anderson}},\ }\href@noop {} {\emph {\bibinfo {title} {An introduction to
  multivariate statistical analysis}}}\ (\bibinfo  {publisher}
  {Wiley-Interscience},\ \bibinfo {address} {Hoboken, N.J},\ \bibinfo {year}
  {2003})\BibitemShut {NoStop}%
\bibitem [{\citenamefont {Neyman}, \citenamefont {Pearson},\ and\ \citenamefont
  {Pearson}(1933)}]{Neyman1933}%
  \BibitemOpen
  \bibfield  {author} {\bibinfo {author} {\bibfnamefont {J.}~\bibnamefont
  {Neyman}}, \bibinfo {author} {\bibfnamefont {E.~S.}\ \bibnamefont {Pearson}},
  \ and\ \bibinfo {author} {\bibfnamefont {K.}~\bibnamefont {Pearson}},\ }\href
  {\doibase 10.1098/rsta.1933.0009} {\bibfield  {journal} {\bibinfo  {journal}
  {Philosophical Transactions of the Royal Society of London. Series A,
  Containing Papers of a Mathematical or Physical Character}\ }\textbf
  {\bibinfo {volume} {231}},\ \bibinfo {pages} {289} (\bibinfo {year}
  {1933})}\BibitemShut {NoStop}%
\bibitem [{\citenamefont {Wilks}(1938)}]{Wilks1938}%
  \BibitemOpen
  \bibfield  {author} {\bibinfo {author} {\bibfnamefont {S.~S.}\ \bibnamefont
  {Wilks}},\ }\href {\doibase 10.1214/aoms/1177732360} {\bibfield  {journal}
  {\bibinfo  {journal} {Ann. Math. Statist.}\ }\textbf {\bibinfo {volume}
  {9}},\ \bibinfo {pages} {60} (\bibinfo {year} {1938})}\BibitemShut {NoStop}%
\bibitem [{\citenamefont {Borisov}(1998)}]{BORISOV1998}%
  \BibitemOpen
  \bibfield  {author} {\bibinfo {author} {\bibfnamefont {G.}~\bibnamefont
  {Borisov}},\ }\href {\doibase https://doi.org/10.1016/S0168-9002(98)00777-3}
  {\bibfield  {journal} {\bibinfo  {journal} {Nuclear Instruments and Methods
  in Physics Research Section A: Accelerators, Spectrometers, Detectors and
  Associated Equipment}\ }\textbf {\bibinfo {volume} {417}},\ \bibinfo {pages}
  {384 } (\bibinfo {year} {1998})}\BibitemShut {NoStop}%
\bibitem [{\citenamefont {Poikela}\ \emph {et~al.}(2014)\citenamefont
  {Poikela}, \citenamefont {Plosila}, \citenamefont {Westerlund}, \citenamefont
  {Campbell}, \citenamefont {Gaspari}, \citenamefont {Llopart}, \citenamefont
  {Gromov}, \citenamefont {Kluit}, \citenamefont {van Beuzekom}, \citenamefont
  {Zappon} \emph {et~al.}}]{timepix3}%
  \BibitemOpen
  \bibfield  {author} {\bibinfo {author} {\bibfnamefont {T.}~\bibnamefont
  {Poikela}}, \bibinfo {author} {\bibfnamefont {J.}~\bibnamefont {Plosila}},
  \bibinfo {author} {\bibfnamefont {T.}~\bibnamefont {Westerlund}}, \bibinfo
  {author} {\bibfnamefont {M.}~\bibnamefont {Campbell}}, \bibinfo {author}
  {\bibfnamefont {M.~D.}\ \bibnamefont {Gaspari}}, \bibinfo {author}
  {\bibfnamefont {X.}~\bibnamefont {Llopart}}, \bibinfo {author} {\bibfnamefont
  {V.}~\bibnamefont {Gromov}}, \bibinfo {author} {\bibfnamefont
  {R.}~\bibnamefont {Kluit}}, \bibinfo {author} {\bibfnamefont
  {M.}~\bibnamefont {van Beuzekom}}, \bibinfo {author} {\bibfnamefont
  {F.}~\bibnamefont {Zappon}},  \emph {et~al.},\ }\href@noop {} {\bibfield
  {journal} {\bibinfo  {journal} {Journal of instrumentation}\ }\textbf
  {\bibinfo {volume} {9}},\ \bibinfo {pages} {C05013} (\bibinfo {year}
  {2014})}\BibitemShut {NoStop}%
\bibitem [{\citenamefont {Fisher-Levine}\ and\ \citenamefont
  {Nomerotski}(2016)}]{timepixcam}%
  \BibitemOpen
  \bibfield  {author} {\bibinfo {author} {\bibfnamefont {M.}~\bibnamefont
  {Fisher-Levine}}\ and\ \bibinfo {author} {\bibfnamefont {A.}~\bibnamefont
  {Nomerotski}},\ }\href@noop {} {\bibfield  {journal} {\bibinfo  {journal}
  {Journal of Instrumentation}\ }\textbf {\bibinfo {volume} {11}},\ \bibinfo
  {pages} {C03016} (\bibinfo {year} {2016})}\BibitemShut {NoStop}%
\bibitem [{\citenamefont {Nomerotski}\ \emph {et~al.}(2017)\citenamefont
  {Nomerotski}, \citenamefont {Chakaberia}, \citenamefont {Fisher-Levine},
  \citenamefont {Janoska}, \citenamefont {Takacs},\ and\ \citenamefont
  {Tsang}}]{Nomerotski2017}%
  \BibitemOpen
  \bibfield  {author} {\bibinfo {author} {\bibfnamefont {A.}~\bibnamefont
  {Nomerotski}}, \bibinfo {author} {\bibfnamefont {I.}~\bibnamefont
  {Chakaberia}}, \bibinfo {author} {\bibfnamefont {M.}~\bibnamefont
  {Fisher-Levine}}, \bibinfo {author} {\bibfnamefont {Z.}~\bibnamefont
  {Janoska}}, \bibinfo {author} {\bibfnamefont {P.}~\bibnamefont {Takacs}}, \
  and\ \bibinfo {author} {\bibfnamefont {T.}~\bibnamefont {Tsang}},\
  }\href@noop {} {\bibfield  {journal} {\bibinfo  {journal} {Journal of
  Instrumentation}\ }\textbf {\bibinfo {volume} {12}},\ \bibinfo {pages}
  {C01017} (\bibinfo {year} {2017})}\BibitemShut {NoStop}%
\bibitem [{\citenamefont {https://www.photonis.com/product/hi-qe
  photocathodes}()}]{Photonis}%
  \BibitemOpen
  \bibfield  {author} {\bibinfo {author} {\bibnamefont
  {https://www.photonis.com/product/hi-qe photocathodes}},\ }\href@noop {} {\
  }\BibitemShut {NoStop}%
\bibitem [{\citenamefont {Winter}\ \emph {et~al.}(2014)\citenamefont {Winter},
  \citenamefont {King}, \citenamefont {Brouard},\ and\ \citenamefont
  {Vallance}}]{P47}%
  \BibitemOpen
  \bibfield  {author} {\bibinfo {author} {\bibfnamefont {B.}~\bibnamefont
  {Winter}}, \bibinfo {author} {\bibfnamefont {S.~J.}\ \bibnamefont {King}},
  \bibinfo {author} {\bibfnamefont {M.}~\bibnamefont {Brouard}}, \ and\
  \bibinfo {author} {\bibfnamefont {C.}~\bibnamefont {Vallance}},\ }\href
  {\doibase 10.1063/1.4866647} {\bibfield  {journal} {\bibinfo  {journal}
  {Review of Scientific Instruments}\ }\textbf {\bibinfo {volume} {85}},\
  \bibinfo {pages} {023306} (\bibinfo {year} {2014})}\BibitemShut {NoStop}%
\bibitem [{\citenamefont {van~der Heijden}\ \emph {et~al.}(2017)\citenamefont
  {van~der Heijden}, \citenamefont {Visser}, \citenamefont {van Beuzekom},
  \citenamefont {Boterenbrood}, \citenamefont {Kulis}, \citenamefont
  {Munneke},\ and\ \citenamefont {Schreuder}}]{spidr}%
  \BibitemOpen
  \bibfield  {author} {\bibinfo {author} {\bibfnamefont {B.}~\bibnamefont
  {van~der Heijden}}, \bibinfo {author} {\bibfnamefont {J.}~\bibnamefont
  {Visser}}, \bibinfo {author} {\bibfnamefont {M.}~\bibnamefont {van
  Beuzekom}}, \bibinfo {author} {\bibfnamefont {H.}~\bibnamefont
  {Boterenbrood}}, \bibinfo {author} {\bibfnamefont {S.}~\bibnamefont {Kulis}},
  \bibinfo {author} {\bibfnamefont {B.}~\bibnamefont {Munneke}}, \ and\
  \bibinfo {author} {\bibfnamefont {F.}~\bibnamefont {Schreuder}},\ }\href@noop
  {} {\bibfield  {journal} {\bibinfo  {journal} {Journal of instrumentation}\
  }\textbf {\bibinfo {volume} {12}},\ \bibinfo {pages} {C02040} (\bibinfo
  {year} {2017})}\BibitemShut {NoStop}%
\bibitem [{\citenamefont {www.amscins.com}()}]{ASI}%
  \BibitemOpen
  \bibfield  {author} {\bibinfo {author} {\bibnamefont {www.amscins.com}},\
  }\href@noop {} {\ }\BibitemShut {NoStop}%
\bibitem [{\citenamefont {Ianzano}\ \emph {et~al.}(2020)\citenamefont
  {Ianzano}, \citenamefont {Svihra}, \citenamefont {Flament}, \citenamefont
  {Hardy}, \citenamefont {Guodong}, \citenamefont {Nomerotski},\ and\
  \citenamefont {Figueroa}}]{Ianzano2020}%
  \BibitemOpen
  \bibfield  {author} {\bibinfo {author} {\bibfnamefont {C.}~\bibnamefont
  {Ianzano}}, \bibinfo {author} {\bibfnamefont {P.}~\bibnamefont {Svihra}},
  \bibinfo {author} {\bibfnamefont {M.}~\bibnamefont {Flament}}, \bibinfo
  {author} {\bibfnamefont {A.}~\bibnamefont {Hardy}}, \bibinfo {author}
  {\bibfnamefont {C.}~\bibnamefont {Guodong}}, \bibinfo {author} {\bibfnamefont
  {A.}~\bibnamefont {Nomerotski}}, \ and\ \bibinfo {author} {\bibfnamefont
  {E.}~\bibnamefont {Figueroa}},\ }\href {\doibase 10.1038/s41598-020-62020-z}
  {\bibfield  {journal} {\bibinfo  {journal} {Scientific Reports}\ }\textbf
  {\bibinfo {volume} {10}},\ \bibinfo {pages} {10} (\bibinfo {year}
  {2020})}\BibitemShut {NoStop}%
\bibitem [{\citenamefont {Nomerotski}\ \emph
  {et~al.}(2020{\natexlab{a}})\citenamefont {Nomerotski}, \citenamefont
  {Katramatos}, \citenamefont {Stankus}, \citenamefont {Svihra}, \citenamefont
  {Cui}, \citenamefont {Gera}, \citenamefont {Flament},\ and\ \citenamefont
  {Figueroa}}]{Nomerotski2020}%
  \BibitemOpen
  \bibfield  {author} {\bibinfo {author} {\bibfnamefont {A.}~\bibnamefont
  {Nomerotski}}, \bibinfo {author} {\bibfnamefont {D.}~\bibnamefont
  {Katramatos}}, \bibinfo {author} {\bibfnamefont {P.}~\bibnamefont {Stankus}},
  \bibinfo {author} {\bibfnamefont {P.}~\bibnamefont {Svihra}}, \bibinfo
  {author} {\bibfnamefont {G.}~\bibnamefont {Cui}}, \bibinfo {author}
  {\bibfnamefont {S.}~\bibnamefont {Gera}}, \bibinfo {author} {\bibfnamefont
  {M.}~\bibnamefont {Flament}}, \ and\ \bibinfo {author} {\bibfnamefont
  {E.}~\bibnamefont {Figueroa}},\ }\href {\doibase 10.1142/s0219749919410272}
  {\bibfield  {journal} {\bibinfo  {journal} {International Journal of Quantum
  Information}\ }\textbf {\bibinfo {volume} {18}},\ \bibinfo {pages} {1941027}
  (\bibinfo {year} {2020}{\natexlab{a}})}\BibitemShut {NoStop}%
\bibitem [{\citenamefont {Nomerotski}\ \emph
  {et~al.}(2020{\natexlab{b}})\citenamefont {Nomerotski}, \citenamefont
  {Keach}, \citenamefont {Stankus}, \citenamefont {Svihra},\ and\ \citenamefont
  {Vintskevich}}]{Nomerotski2020_1}%
  \BibitemOpen
  \bibfield  {author} {\bibinfo {author} {\bibfnamefont {A.}~\bibnamefont
  {Nomerotski}}, \bibinfo {author} {\bibfnamefont {M.}~\bibnamefont {Keach}},
  \bibinfo {author} {\bibfnamefont {P.}~\bibnamefont {Stankus}}, \bibinfo
  {author} {\bibfnamefont {P.}~\bibnamefont {Svihra}}, \ and\ \bibinfo {author}
  {\bibfnamefont {S.}~\bibnamefont {Vintskevich}},\ }\href {\doibase
  10.3390/s20123475} {\bibfield  {journal} {\bibinfo  {journal} {Sensors}\
  }\textbf {\bibinfo {volume} {20}},\ \bibinfo {pages} {3475} (\bibinfo {year}
  {2020}{\natexlab{b}})}\BibitemShut {NoStop}%
\bibitem [{\citenamefont {Zhao}\ \emph {et~al.}(2017)\citenamefont {Zhao},
  \citenamefont {van Beuzekom}, \citenamefont {Bouwens}, \citenamefont
  {Byelov}, \citenamefont {Chakaberia}, \citenamefont {Cheng}, \citenamefont
  {Maddox}, \citenamefont {Nomerotski}, \citenamefont {Svihra}, \citenamefont
  {Visser} \emph {et~al.}}]{Zhao2017}%
  \BibitemOpen
  \bibfield  {author} {\bibinfo {author} {\bibfnamefont {A.}~\bibnamefont
  {Zhao}}, \bibinfo {author} {\bibfnamefont {M.}~\bibnamefont {van Beuzekom}},
  \bibinfo {author} {\bibfnamefont {B.}~\bibnamefont {Bouwens}}, \bibinfo
  {author} {\bibfnamefont {D.}~\bibnamefont {Byelov}}, \bibinfo {author}
  {\bibfnamefont {I.}~\bibnamefont {Chakaberia}}, \bibinfo {author}
  {\bibfnamefont {C.}~\bibnamefont {Cheng}}, \bibinfo {author} {\bibfnamefont
  {E.}~\bibnamefont {Maddox}}, \bibinfo {author} {\bibfnamefont
  {A.}~\bibnamefont {Nomerotski}}, \bibinfo {author} {\bibfnamefont
  {P.}~\bibnamefont {Svihra}}, \bibinfo {author} {\bibfnamefont
  {J.}~\bibnamefont {Visser}},  \emph {et~al.},\ }\href {\doibase
  10.1063/1.4996888} {\bibfield  {journal} {\bibinfo  {journal} {Review of
  Scientific Instruments}\ }\textbf {\bibinfo {volume} {88}},\ \bibinfo {pages}
  {113104} (\bibinfo {year} {2017})}\BibitemShut {NoStop}%
\end{thebibliography}%

\end{document}